\documentclass[particles,article,accept,moreauthors,pdftex,10pt,a4paper,particles]{Definitions/mdpi} 
\usepackage{amssymb,color,graphicx,bm,mathrsfs,color,amsmath,slashed,comment}
\newcommand{\bea}{\begin{eqnarray}}
\newcommand{\eea}{\end{eqnarray}}
\newcommand{\be}{\begin{equation}}
\newcommand{\ee}{\end{equation}}

\newcommand{\vecp}{{\bm p}}

\newcommand{\Tr}{{\rm Tr}}

\newcommand{\ie}{{i.e.}}
\newcommand{\eg}{{e.g.}}
\input{acronym.input}
\definecolor{red}{rgb}{0.8,0,0}
\definecolor{violet}{rgb}{0.4,0,0.4}
\definecolor{green}{rgb}{0,0.5,0.0}
\definecolor{navy}{rgb}{0.0,0.0,0.6}
\definecolor{orange}{rgb}{0.8,0.2,0.0}

\usepackage[normalem]{ulem}  

\firstpage{1} 
\articlenumber{212-229}
\doinum{10.3390/particles1010016 }
\pubvolume{1}
\issuenum{1}
\pubyear{2018}
\copyrightyear{2018}
\history{Received: 18 June 2018; Accepted: 13 August 2018; Published: 20 August 2018}
\updates{yes}

\Title{ Bulk Viscosity of Hot Quark Plasma from Non-Equilibrium
  Statistical Operator}

\Author{
 Arus Harutyunyan  $^{1,}$*\orcidA{}
and   Armen Sedrakian    $^{2}$\orcidB{} 
            }

\address{\noindent 
$^{1}$ \quad \textls[-25]{Institute for Theoretical Physics, Goethe 
  University, Max-von-Laue-Stra\ss e, 1,~60438~Frankfurt~am~Main,~Germany} \\
  
$^{2}$ \quad  Frankfurt Institute for Advanced Studies,
Ruth-Moufang-Stra\ss e, 1, 60438 Frankfurt am Main,
  Germany; sedrakian@fias.uni-frankfurt.de}
               
\corres{Correspondence: arus@th.physik.uni-frankfurt.de; Tel.:+49 069 798 47691}
               
\AuthorNames{Arus Harutyunyan, Armen Sedrakian}

\abstract{ We provide a discussion of the bulk viscosity of two-flavor
  quark plasma, described by the Nambu--Jona-Lasinio model, within the
  framework of Kubo-Zubarev formalism. This discussion, which is
  complementary to our earlier study,
  contains a new, detailed derivation of the bulk viscosity in the
  case of multiple conserved charges. We also provide some numerical
  details of the computation of the bulk viscosity close to the Mott
  transition line, where the dissipation is dominated by decays of
  mesons into quarks and their inverse processes. We close with a
  summary of our current understanding of this quantity, which
  stresses the importance of loop resummation for obtaining the
  qualitatively correct result near the Mott line. }


\begin{document}

\section{Introduction}
\label{sec:hydro_intro}

Transport coefficients of hot and dense quark plasma are key inputs in
the hydrodynamical description of the heavy-ion experiments at
Relativistic Heavy Ion Collider (RHIC) and Large Hadron Collider
(LHC).  The matter created in these experiments exhibits a very small
ratio of the shear viscosity to the entropy density, which is close to the lower
bound placed by the uncertainty principle~\cite{Danielewicz:1985} and
conjectured on the basis of AdS/CFT duality~\cite{Kovtun:2005}.

The bulk viscosity describes the dissipation in cases where
pressure falls out of its equilibrium value on uniform expansion or
contraction of fluid.  It vanishes in several cases, \eg, for an
ultrarelativistic or nonrelativistic gas interacting weakly with local
forces via binary collisions, \mbox{as well as} in strongly coupled systems
with conformal symmetry.  Because at high energies QCD is almost
conformally symmetric, the bulk viscosity of quark-gluon
plasma is small in the perturbative
regime~\cite{Hosoya:1985,Arnold:2006,Moore:2008,Chen:2013}.  At low
energies, the conformal symmetry is broken by the quark mass 
and/or by dimensionful regularization of the ultraviolet divergences,
in which case the bulk viscous effects can become important.  Large
values of bulk viscosity were found, in particular, close to the
chiral phase transition line~\cite{Meyer:2008,Paech:2006}.
Computations of the QCD bulk viscosity in the strongly coupled regime
where carried out using various methods including lattice
simulations~\cite{Moore:2008,Karsch:2008,Aarts:2007}, quasiparticle
Boltzmann transport~\cite{Sasaki:2010,Chakraborty:2011,
  Chandra:2011,Dobado:2012,Marty:2013} and the Kubo
formalism~\cite{Harutyunyan:2017b,Xiao:2014,Ghosh:2016}.

The focus of this contribution, which is complementary to our earlier
study~\cite{Harutyunyan:2017b}, is the bulk viscosity of quark matter
in the non-perturbative regime as it is realized close to the chiral
phase transition line.  The case of bulk viscosity is special because
it requires a resummation of an infinite series of loop diagrams,
whereas the remaining coefficient (\mbox{shear viscosity}, thermal and
electrical conductivities) are given by the one-loop result only, see
Ref.~\cite{Harutyunyan:2017a}.  Specifically, our aim here is to
provide details of the computation of this quantity which complement
our earlier publication~\cite{Harutyunyan:2017b}. First, we provide a
formal derivation of the bulk viscosity coefficient from the Zubarev
formalism of non-equilibrium statistical operator
(NESO)~\cite{Zubarev:1974,Zubarev:1997} in a general setting of
relativistic quantum field theory assuming a system with multiple
conserved charges. We then go on to discuss the details of
diagrammatic evaluation of the bulk viscosity within the
Nambu--Jona-Lasinio (NJL) model, which is an effective field theory of
QCD that captures its chiral symmetry breaking feature.  The main
mechanism of dissipation within this model is provided by the mesonic
fluctuations close to the critical line of chiral phase transition.

Another important ingredient of the diagrammatic evaluation of the
two-point correlation function determining the bulk viscosity is the
$1/N_c$ expansion~\cite{Quack:1994}.  Several recent computations
of bulk viscosity, which were based on a Kubo formula and the NJL
model, evaluated the relevant correlation function at the one-loop
level~\cite{Xiao:2014,Ghosh:2016}.  However, our recent
analysis~\cite{Harutyunyan:2017b} indicates, that the one-loop
approximation is not consistent with the $1/N_c$ power counting scheme
and a resummation of infinite series of loop diagrams is required to
obtain the leading-order approximation to the bulk viscosity. Below,
we provide some numerical details of this computation.

This work is organized as follows. In Section~\ref{sec:Zub_formalism} we
review Zubarev's method of the NESO and derive a Kubo-type formula for
the bulk viscosity.  Section~\ref{sec:Kubo_bulk} is devoted to the
application of the general formalism to the case of two-flavor quark matter
described by the NJL model. Our numerical results for the bulk
viscosity are collected in Section~\ref{sec:results}.
Section~\ref{sec:conclusions} provides a short summary.  \mbox{The
computation} of a Matsubara sum, which is used in the derivation of the
Kubo formula, is relegated to  \mbox{Appendix}~\ref{app:sums}.  We use the natural
(Gaussian) units with $\hbar= c=k_B=1$, and the metric signature
$g^{\mu\nu}={\rm diag}(1,-1,-1,-1)$.

\section{Bulk Viscosity Formula from the Non-Equilibrium Statistical
  Operator}
\label{sec:Zub_formalism}

The coefficient of the bulk viscosity was computed within the NESO method
for the case of a system without conserved charges in the seminal
paper by Hosoya et al.~\cite{Hosoya:1984}. Our purpose here is to
extend that derivation to the case of systems with multiple conserved
charges.

Hydrodynamics of relativistic quantum fluids is described by the
energy-momentum tensor $\hat{T}^{\mu\nu}(x)$ and currents of conserved
charges $\hat{N}^{\mu}_a(x)$. We consider the general case of multiple
conserved charge flavors (\eg, baryonic, electric, etc.) which are
labeled by the index $a$. In this case, these conservation laws take
the form
\be\label{eq:cons_laws}
\partial_{\mu} \hat{T}^{\mu\nu}(x) =0,\qquad
\partial_{\mu} \hat{N}_a^{\mu}(x) =0.
\ee

For systems in the hydrodynamic regime, one can introduce local
thermodynamic variables, such as temperature
$T(x)\equiv \beta^{-1}(x)$, chemical potentials $\mu_a(x)$ and fluid
4-velocity $u^{\nu}(x)$ as smooth functions of the space-time
coordinates $x\equiv (\bm x,t)$. Below we will specify the (matching)
conditions which are necessary to identify these quantities.  Our next
step is to define a NESO as
\be\label{eq:stat_op_full}
\hat{\rho}(t) = Q^{-1}e^{-\hat{A}+\hat{B}},
\qquad Q=\Tr e^{-\hat{A}+\hat{B}},
\ee
with operators defined as
\bea\label{eq:A_op}
\hat{A}(t)&=&\int d^3x \Big[\beta^\nu(x)
\hat{T}_{0\nu}(x)-
\sum_a \alpha_a(x) \hat{N}_a^0(x)\Big],\\
\label{eq:B_op}
\hat{B}(t)&=& \lim_{\varepsilon\to +0}\int d^3x_1 
\int_{-\infty}^tdt_1 e^{\varepsilon(t_1-t)} \hat{C}(x_1),\\
\label{eq:C_op}
\hat{C}(x)&=&\hat{T}_{\mu\nu}(x)
\partial^{\mu}\beta^\nu(x)-
\sum_a\hat{N}_a^\mu(x)\partial_\mu\alpha_a(x),
\eea
where the covariant quantities ($c$-numbers) are defined as
\be\label{eq:beta_nu_alpha}
\beta^\nu(x) = \beta(x) u^{\nu}(x),\qquad
\alpha_a (x)=\beta(x)\mu_a (x).
\ee

Note that the limit $\varepsilon\to +0$ in Equation~\eqref{eq:B_op} should
be taken only after the thermodynamic limit.  The proper order of
taking the thermodynamic and $\varepsilon\to +0$ limits guarantees
that the NESO in Equation~\eqref{eq:stat_op_full} satisfies the Liouville
equation with an infinitesimal source term, which breaks the
time-reversibility of that equation and chooses its retarded solution
for positive values of $\varepsilon$~\cite{Hosoya:1984,Zubarev:1974,
  Zubarev:1997}.
Equations~\eqref{eq:stat_op_full}--\eqref{eq:beta_nu_alpha} 
generalize the analogous expressions of Refs.~\cite{Hosoya:1984,Huang:2011} to the
case of a system with multiple conserved charges.

The first term in the exponent in Equation~\eqref{eq:stat_op_full} corresponds 
to the local equilibrium part of the statistical operator, defined as
\be\label{eq:stat_op_local}
\hat{\rho}_l(t) = Q^{-1}_le^{-\hat{A}},
\qquad Q_l=\Tr e^{-\hat{A}}.
\ee

The second term in the exponent of Equation~\eqref{eq:stat_op_full} is the
non-equilibrium part of the statistical operator, which can be
interpreted as a thermodynamic ``force''. For small deviations from
equilibrium, it can be treated as a small perturbation.  Expanding the
NESO around the local equilibrium distribution and keeping linear in
the operator $\hat{B}$ terms we find
\bea  \label{eq:stat_approx}
\hat{\rho} = \left[1+
\int_0^1d\tau \left(e^{-\tau \hat{A}}\hat{B}e^{\tau \hat{A}}
 - \langle \hat{B}\rangle_l\right)\right]\hat{\rho}_{l}.
\eea

The statistical average of any operator
$\hat{X}(x)$ can be written now as 
\bea\label{eq:stat_average}
\langle \hat{X}(x)\rangle 
= \Tr[\hat{\rho}(t)\hat{X}(x)]=
\langle \hat{X}(x)\rangle_l +
\int d^3x_1\int_{-\infty}^tdt_1 e^{\varepsilon(t_1-t)}
\Big(\hat{X}(x),\hat{C}(x_1)\Big),
\eea
where $\langle \hat{X}(x)\rangle_l$ is the local equilibrium average
and a two-point correlation function has been defined
as~\cite{Hosoya:1984,Huang:2011}
\bea\label{eq:2_point_corr}
\Big(\hat{X}(x),\hat{C}(x_1)\Big) =
\int_0^1 d\tau \langle\hat{X}(x)
\left[e^{-\tau \hat{A}}\hat{C}(x_1)e^{\tau \hat{A}}
 - \langle \hat{C}(x_1)\rangle_l\right]\rangle_l.
\eea

The final point of our general discussion of the NESO method is the
procedure by which the quantities $\beta^\nu$ and $\alpha_a$ are
matched with the relevant thermodynamic variables in an arbitrary
non-equilibrium state.  This can be achieved by the following 
{\it matching conditions}~\cite{Zubarev:1974,Zubarev:1997,Huang:2011}
\bea\label{eq:matching}
\langle \hat{\epsilon}(x)\rangle
=\langle\hat{\epsilon}(x)\rangle_l,\qquad
\langle\hat{n}_a(x)\rangle
=\langle\hat{n}_a(x)\rangle_l,
\eea
where the operators of the energy and charge densities are defined as
$\hat{\epsilon}=u_\mu u_\nu \hat{T}^{\mu\nu}$ and
$\hat{n}_a=u_\mu \hat{N}_a^\mu$.  \mbox{In these expressions}, the fluid
4-velocity $u^\mu$, which is normalized to unity $u_\mu u^\mu=1$,
should be ``tied'' to a physical current.  This could be either the
energy flow, which specifies the Landau-Lifshitz
frame~\cite{Landau:1987} or the charge flow, which specifies the
Eckart frame~\cite{Eckart:1940}.  In the Landau frame
\mbox{$u_\mu\langle\hat{T}^{\mu\nu}\rangle = \langle\hat{\epsilon}\rangle
u^\nu$},
whereas in the Eckart frame
$\langle\hat{N}_a^\mu\rangle = \langle\hat{n}_a\rangle u^\mu$.  The
conditions \eqref{eq:matching} define the temperature and the chemical
potentials of components as {\it non-local functionals} of
$\langle \hat{\epsilon}(x)\rangle$ and
$\langle\hat{n}_a(x)\rangle$~\cite{Zubarev:1972}. However, \mbox{the
hydrodynamic} description requires thermodynamic parameters as {\it
  local} functions of the energy and charge densities. In practice,
this difficulty is circumvented by dividing the fluid into elements
which are in local statistical equilibrium and each of which is
independent of the other~\cite{Mori:1958}. In practice, \mbox{the local}
equilibrium values $\langle\hat{\epsilon}\rangle_l$ and
$\langle\hat{n}_a\rangle_l$ in Equation~\eqref{eq:matching} are then
evaluated assuming formally {\it constant values} of $\beta$ and
$\mu_a$, which are identified by matching
$\langle\hat{\epsilon}\rangle_l$ and $\langle \hat{n}_a\rangle_l$ to
the real values of these quantities $\langle\hat{\epsilon}\rangle$ and
$\langle \hat{n}_a\rangle$ at any given point $x$. In this way one can construct a {\it fictitious local equilibrium state},
characterized by the thermodynamic parameters $\beta(x)$ and
$\mu_a(x)$, such that it reproduces the local values of the energy and
charge densities at every point of the space and time.

\subsection{Decomposition into Different Dissipative Processes}
\label{sec:decomp}
 
\textls[-20]{To identify the different dissipative processes, we now
exploit the common decompositions of the energy-momentum tensor and
the charge currents into the ideal and dissipative parts}
\bea \label{eq:T_munu_decomp}
\hat{T}^{\mu\nu} &=& \hat{\epsilon} u^{\mu}u^{\nu} - \hat{p}\Delta^{\mu\nu} + \hat{q}^{\mu}u^{\nu}+ \hat{q}^{\nu}u^{\mu} + \hat{\pi}^{\mu\nu},\\
\label{eq:N_decomp}
\hat{N}^{\mu}_a &=& \hat{n}_au^\mu +\hat{j}_a^{\mu},
\eea
where $\hat{p}$ is the operator of pressure; $\Delta^{\mu\nu}=g^{\mu\nu}-u^\mu u^\nu$ is the projection operator onto the 3-space orthogonal to $u_\mu$
and has the properties
\bea\label{eq:prop_proj}
u_\mu\Delta^{\mu\nu}=\Delta^{\mu\nu}u_\nu=0,\qquad
\Delta^{\mu\nu}\Delta_{\nu\lambda}=\Delta^{\mu}_\lambda,\qquad \Delta^{\mu}_{\mu}=3.
\eea 

The dissipative quantities $\hat{\pi}^{\mu\nu}$, $\hat{q}^\mu$ and $\hat{j}^\mu_a$ are the operators of the shear stress tensor, energy diffusion flux and charge diffusion fluxes, respectively, and they satisfy the following conditions 
\bea \label{eq:orthogonality}
\quad
u_{\nu}\hat{q}^{\nu} = 0,\qquad
u_{\nu}\hat{j}_a^{\nu} = 0,\qquad 
u_{\nu}\hat{\pi}^{\mu\nu} = 0,\qquad 
\hat{\pi}_{\mu}^\mu=0.
\eea

The operators on the right-hand sides of Equations~\eqref{eq:T_munu_decomp} and \eqref{eq:N_decomp} can be obtained via the projections of
 $\hat{T}^{\mu\nu}$ and $\hat{N}^\mu_a$
\bea\label{eq:proj1_op}
\hat{\epsilon} = u_\mu u_\nu \hat{T}^{\mu\nu},\qquad
\hat{n}_a = u_\mu\hat{N}^{\mu}_a,\qquad
\hat{p}=-\frac{1}{3}\Delta_{\mu\nu}
\hat{T}^{\mu\nu},\\
\label{eq:proj2_op}
\hat{\pi}^{\mu\nu} = \Delta_{\alpha\beta}^{\mu\nu} \hat{T}^{\alpha\beta},\qquad
\hat{q}^\mu  = u_\alpha\Delta_{\beta}^{\mu}\hat{T}^{\alpha\beta},\qquad
\hat{j}^{\nu}_a=\Delta_{\mu}^{\nu} \hat{N}_a^{\mu},
\eea
which in the local rest frame [$u_\mu=(1,0,0,0)$] read
\bea \label{eq:currents_rest1_op}
\hat{\epsilon} = \hat{T}^{00},\qquad
\hat{n}_a = \hat{N}^{0}_a,\qquad
\hat{p} =-\frac{1}{3}\hat{T}^k_k,\hspace{1.5cm}\\
\label{eq:currents_rest2_op}
\hat{\pi}_{kl} = \left(\delta_{ki}\delta_{lj}-\frac{1}{3}\delta_{kl}\delta_{ij}
\right) \hat{T}_{ij},\qquad
\hat{q}^i  = \hat{T}^{0i},\qquad
\hat{j}_a^{i} = \hat{N}^{i}_a.
\eea
In Equation~\eqref{eq:proj2_op} we introduced a fourth-rank traceless
projector orthogonal to $u^\mu$
\bea\label{eq:projector_delta4}
\Delta_{\mu\nu\rho\sigma}= \frac{1}{2}\left(\Delta_{\mu\rho}\Delta_{\nu\sigma}
+\Delta_{\mu\sigma}\Delta_{\nu\rho}\right)
-\frac{1}{3}\Delta_{\mu\nu}\Delta_{\rho\sigma}.
\eea

The hydrodynamic quantities ${\pi}^{\mu\nu}$, ${q}^\mu$ and ${j}^\mu_a$ are obtained as the statistical averages of the corresponding operators over the NESO according to Equations~\eqref{eq:stat_average} and \eqref{eq:2_point_corr}. In local equilibrium the averages of the dissipative operators vanish:
\bea\label{eq:diss_currents_av_eq}
\langle \hat{q}^{\mu}\rangle_l =0,\qquad
\langle \hat{j}^{\mu}\rangle_l =0,\qquad
\langle \hat{\pi}^{\mu\nu}\rangle_l =0.
\eea

To compute the non-equilibrium averages of these operators it is
convenient to write the operator $\hat{C}$ given by
Equation~\eqref{eq:C_op} as a sum of contributions of different dissipative processes according to
Equations~\eqref{eq:T_munu_decomp} and \eqref{eq:N_decomp}.  Similar
decompositions were performed in
Refs.~\cite{Hosoya:1984,Huang:2011}.
Recalling the properties \eqref{eq:prop_proj} and
\eqref{eq:orthogonality} we obtain
\bea\label{eq:op_C_decompose1}
\hat{C}
=\hat{\epsilon} D\beta - \hat{p}
\beta\theta -\sum\limits_a\hat{n}_aD\alpha_a + \hat{q}^{\lambda}(\beta Du_{\lambda}+\nabla_{\lambda}\beta)
-\sum\limits_a\hat{j}^{\lambda}_a \nabla_\lambda\alpha_a
+ \beta\hat{\pi}_{\lambda\rho}\sigma^{\lambda\rho},
\eea
where we introduced the covariant time-derivative
$D = u^{\rho}\partial_{\rho}$, the covariant spatial derivative
\mbox{$\nabla_\rho= \Delta_{\rho\lambda}\partial^\lambda$}, the fluid
expansion rate $\theta = \partial_{\rho}u^{\rho}$ and the velocity
stress tensor via
$\sigma^{\lambda\rho}=\Delta^{\lambda\rho\alpha\beta}\partial_\alpha
u_\beta$.
As a first approximation, we can eliminate the terms $D\beta$,
$D\alpha_a$ and $Du_\lambda$ using the equations of ideal 
hydrodynamics
\bea \label{eq:ideal_hydro}
Dn_a+ n_a\theta =0,\qquad
D\epsilon+ h \theta =0,\qquad
h D u_\lambda = \nabla_\lambda p,
\eea
where $p=p(\epsilon, n_a)$ is the equilibrium pressure fixed 
by an equation of state, and $h=\epsilon+p$ is the enthalpy density.
Choosing $\epsilon$ and $n_a$
 as independent thermodynamic variables and using the first two 
equations in \eqref{eq:ideal_hydro} we can write
\bea\label{eq:D_beta}
D\beta &=& \left(\frac{\partial\beta}{\partial\epsilon}\right)_{n_a}D\epsilon +
\sum\limits_a
\left(\frac{\partial\beta}{\partial n_a}
\right)_{\epsilon, n_b\neq n_a}Dn_a
=-h\theta \left(\frac{\partial\beta}{\partial\epsilon}\right)_{n_a} -\sum\limits_a
n_a\theta \left(\frac{\partial\beta}{\partial n_a}
\right)_{\epsilon, n_b\neq n_a},\\
\label{eq:D_alpha}
D\alpha_c &=& \left(\frac{\partial\alpha_c}{\partial\epsilon}\right)_{n_a}D\epsilon +
\sum\limits_a
\left(\frac{\partial\alpha_c}{\partial n_a}
\right)_{\epsilon, n_b\neq n_a}Dn_a
= -h\theta \left(\frac{\partial\alpha_c}{\partial\epsilon}\right)_{n_a} -\sum\limits_a
n_a\theta \left(\frac{\partial\alpha_c}{\partial n_a}
\right)_{\epsilon, n_b\neq n_a}\!.\qquad
\eea

In the next step we exploit the thermodynamic relations
\bea\label{eq:thermodyn6}
ds=\beta d\epsilon -\sum\limits_a\alpha_a dn_a,\qquad
\beta dp  =-hd\beta +\sum\limits_a n_ad\alpha_a,
\eea
to obtain
\bea\label{eq:rel_alpha_beta}
\left(\frac{\partial \beta}{\partial n_a}
\right)_{\epsilon,n_b\neq n_a}=-
\left(\frac{\partial \alpha_a}{\partial\epsilon}
\right)_{n_b},\qquad
\left(\frac{\partial \alpha_c}{\partial n_a}
\right)_{\epsilon, n_b\neq n_a}=
\left(\frac{\partial \alpha_a}{\partial n_c}
\right)_{\epsilon, n_b\neq n_c},\\
\label{eq:h_n_a}
h=-\beta\left(\frac{\partial p}{\partial\beta}
\right)_{\alpha_a},\qquad
n_a=\beta\left(\frac{\partial p}{\partial\alpha_a}
\right)_{\beta,\alpha_b\neq \alpha_a}.\hspace{2cm}
\eea
Substituting these relations back into Equations~\eqref{eq:D_beta} and  \eqref{eq:D_alpha} we obtain
\bea\label{eq:D_beta_alpha}
D\beta =
\beta\theta \left(\frac{\partial p}{\partial\epsilon}
\right)_{n_a},\qquad
D\alpha_c =  -\beta\theta\left(\frac{\partial p}{\partial n_c} \right)_{\epsilon, n_b\neq n_c}.
\eea
Now the first three terms in Equation~\eqref{eq:op_C_decompose1}
can be combined as follows
\bea\label{eq:scalar_part}
\hat{\epsilon} D\beta- \hat{p}
\beta\theta -\sum\limits_a\hat{n}_a D\alpha_a =
- \beta\theta \hat{p}^*,
\eea
where
\bea\label{eq:p_star}
\hat{p}^* 
= \hat{p} - \left(\frac{\partial p}{\partial\epsilon}
\right)_{n_a}\hat{\epsilon}
-\sum\limits_a\left(\frac{\partial p}{\partial n_a} \right)_{\epsilon, n_b\neq n_a} \hat{n}_a.
\eea

\subsection{Kubo Formula for the Bulk Viscosity}
\label{sec:diss1}

By definition, bulk viscous pressure $\Pi$ measures the deviation of
the thermodynamic pressure from its equilibrium value, which results
from the expansion or compression of the fluid. Therefore, \mbox{it might}
appear at a first glance that the bulk viscous pressure should be
identified as $\langle\hat{p}\rangle-\langle\hat{p}\rangle_l$.
However, it is easy to see that such a definition would be erroneous.
To understand the problem, we go back to the matching
conditions \eqref{eq:matching}, which define the local equilibrium
state. \mbox{As explained above}, these conditions
are satisfied only if the local equilibrium distribution function is evaluated
formally assuming {\it uniform} background values of the thermodynamic
parameters, \ie, as if these were constant in space and time with the
given values $\beta(x)$ and $\mu_a(x)$. Because the local equilibrium
distribution \eqref{eq:stat_op_local} is actually a functional of {\it
  non-uniform} thermodynamic parameters, \mbox{the average} values
$\langle\hat{\epsilon}\rangle_l$ and $\langle\hat{n}_a\rangle_l$ in
the full computation are shifted from the actual values of
$\epsilon=\langle\hat\epsilon\rangle$ and
$n_a=\langle\hat{n}_a\rangle$ by additional gradient terms
$\Delta\epsilon \equiv
\langle\hat{\epsilon}\rangle-\langle\hat{\epsilon}\rangle_l$
and
$\Delta n_a \equiv \langle\hat{n}_a\rangle-\langle\hat{n}_a\rangle_l$,
which were neglected in Equation~\eqref{eq:matching}. These shifts bring,
in their turn, an additional shift in the equilibrium part of the
pressure $\langle\hat{p}\rangle_l$, which should not be included in
the bulk viscous pressure~\cite{Arnold:2006, Dusling:2012,Jeon:1995}.  Thus, \mbox{the
bulk} viscous pressure should be defined as the difference between the
actual non-equilibrium pressure $\langle\hat{p}\rangle$ and the
equilibrium pressure ${p}(\epsilon,n_a)$, which is {\it not equal} to
$\langle\hat{p}\rangle_l\equiv
{p}(\langle\hat{\epsilon}\rangle_l,\langle\hat{n}_a\rangle_l)$:
\bea\label{eq:pressure_expand}
 p({\epsilon},{n}_a)=
{p}(\langle\hat{\epsilon}\rangle_l+\Delta\epsilon,\langle\hat{n}_a\rangle_l +\Delta n_a)
=\langle\hat{p}\rangle_l +\left(\frac{\partial p}{\partial\epsilon}
\right)_{n_a}\Delta\epsilon +\sum\limits_a\left(\frac{\partial p}{\partial n_a} \right)_{\epsilon, n_b\neq n_a}\Delta n_a,
\eea
where we kept only the linear terms.  Then, the bulk
viscous pressure is given by 
\bea\label{eq:bulk_1} \Pi \equiv \langle\hat{p}\rangle
-{p}(\epsilon,n_a)=\langle\hat{p}^*\rangle -\langle\hat{p}^*\rangle_l,
\eea
where we used the definition of $\hat{p}^*$ given by Equation~\eqref{eq:p_star}.
From Equations~\eqref{eq:stat_average}, \eqref{eq:op_C_decompose1}
and \eqref{eq:scalar_part} we \mbox{then obtain }
\bea\label{eq:pressure_av} \Pi =- \beta\theta \int d^3x_1
\int_{-\infty}^tdt_1 e^{\varepsilon(t_1-t)} \Big(\hat{p}^*
(x),\hat{p}^*(x_1)\Big), \eea
where we dropped the correlators between operators of different rank, 
because they vanish in isotropic medium according to Curie's
theorem~\cite{deGroot:1963}. Introducing the bulk viscosity as
\bea \label{eq:bulk_def}
\zeta = \beta \int d^3x_1
\int_{-\infty}^tdt_1 e^{\varepsilon(t_1-t)}
\Big(\hat{p}^*(x), \hat{p}^*(x_1)\Big),
\eea
we rewrite Equation~\eqref{eq:pressure_av} as
\bea \label{eq:shear_bulk_1}
\Pi =-\zeta\theta.
\eea

The correlator \eqref{eq:bulk_def} can be evaluated using uniform
background values of thermodynamic parameters, \ie, as if the system
is in {\it global thermal equilibrium}.  Finally, the bulk viscosity
can be cast in the form of a Kubo
formula~\cite{Hosoya:1984,Huang:2011}
\bea \label{eq:bulk_kubo}
\zeta = -\frac{d}{d\omega} {\rm Im}G^R_{\zeta}(\omega)\bigg\vert_{\omega=0},
\eea
where the two-point retarded {\it equilibrium} Green's function in the 
zero-wavenumber limit is given by 
\bea\label{eq:green_func}
G^R_{\zeta}(\omega)= -i\int_{0}^{\infty} dt
e^{i\omega t}\int d^3x\langle\big[{\hat{p}^{*}}(\bm x, t),
{\hat{p}^{*}}(\bm 0,0)\big]\rangle_l,
\eea
where the square brackets denote a commutator.

\section{Bulk Viscosity within the Two-Flavor NJL Model}
\label{sec:Kubo_bulk}

In this section, we illustrate the computation of the bulk viscosity
following Ref.~\cite{Harutyunyan:2017b}; in doing so we will provide
some numerical details not exposed earlier.  The Lagrangian of the
two-flavor NJL model contains scalar-isoscalar and
pseudoscalar-isovector channels of interactions among quarks and is
given by
\be\label{eq:lagrangian}
\mathcal{L}=\bar\psi(i\slashed \partial-m_0)\psi+
\frac{G}{2}\left[(\bar\psi\psi)^2+
(\bar\psi i\gamma_5\bm\tau\psi)^2\right],
\ee 
where $\psi=(u,d)^T$ is the Dirac field for $u$ and $d$ quarks,
$m_0=5.5$ MeV is the current-quark mass, \mbox{$G=10.1$ GeV$^{-2}$} is the
effective coupling and $\bm\tau$ is the vector of Pauli matrices in
the space of isospin.  \mbox{The NJL model} is regularized with a
three-momentum cut-off $\Lambda = 0.65$ GeV. Assuming
isospin-symmetry, the only conserved current is the net particle
current given by
\bea\label{eq:current}
\hat{N}_\mu=\bar{\psi}\gamma_\mu\psi.
\eea

 The energy-momentum tensor reads
\be\label{eq:energymom}
\hat{T}_{\mu\nu}=\frac{i}{2}(\bar\psi\gamma_{\mu}
\partial_{\nu}\psi +\bar\psi\gamma_{\nu}
\partial_{\mu}\psi)-g_{\mu\nu}\mathcal{L}.
\ee 

The relevant operator $\hat{p}^*$ which enters the Kubo
formula~\eqref{eq:bulk_kubo} with the correlator given by
Equation~\eqref{eq:green_func} in the local rest frame reads (see
Equations~\eqref{eq:currents_rest1_op} and \eqref{eq:p_star})
\bea\label{eq:p_star1}
\hat{p}^*
=\frac{1}{3}\hat{T}_{ii} -\bigg(\frac{\partial p}
{\partial\epsilon}\bigg)_n \hat{T}_{00}-\bigg(\frac{\partial p}
{\partial n}\bigg)_\epsilon \hat{N}_0.
\eea

\begin{figure}[H] 
\begin{center}
\vspace{-0.4cm}
\includegraphics[width=5cm,keepaspectratio]{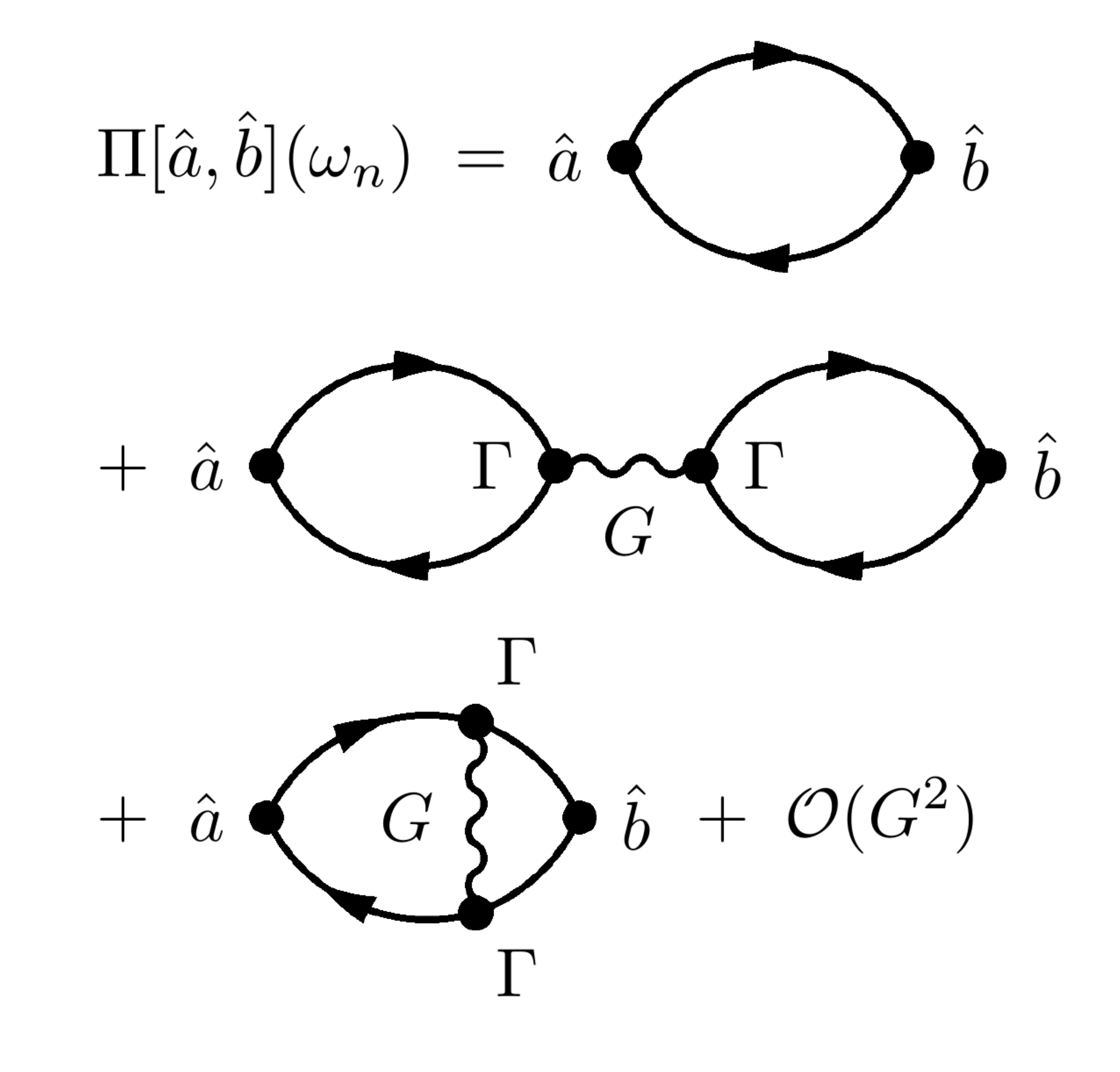}
\vspace{-0.6cm}
\caption{ Contributions to the two-point correlation functions from
  ${\cal O}(N_c^1)$ (first and second lines) and ${\cal O}(N_c^0)$
  (the third line) diagrams which are either of zeroth or first order
  in the coupling constant $G$.  The interaction vertex 
  $\Gamma$ stands for the strong (scalar or pseudoscalar) vertex. The
operators $\hat a$ and $\hat b$ are defined in the text.}
\label{fig:loops_ab} 
\end{center}
\end{figure}

Inserting Equation~\eqref{eq:p_star1} back into Equation~\eqref{eq:green_func} and using the
symmetry relations $[\hat{T}_{11},\hat{T}_{22}]=[\hat{T}_{22},\hat{T}_{11}]=[\hat{T}_{11},\hat{T}_{33}]$,
etc., we obtain
\begin{equation}
\begin{array}{lll}
\label{eq:corp2} 
G^{R}_\zeta(\omega) &=& -i\int_{0}^{\infty} dt
e^{i\omega t}\int d^3 x\langle
\frac{1}{3}[\hat{T}_{11},\hat{T}_{11}]+\frac{2}{3}[\hat{T}_{11},\hat{T}_{22}]
-2\gamma[\hat{T}_{11},\hat{T}_{00}]\\[1ex]
&&-2\delta[\hat{T}_{11},\hat{N}_{0}]+ 2\gamma\delta
[\hat{T}_{00},\hat{N}_{0}]+\gamma^2[\hat{T}_{00},\hat{T}_{00}]
+\delta^2[\hat{N}_{0},\hat{N}_{0}]\rangle_l,
\end{array}
\end{equation}
where we omitted the arguments of the operators.
Substituting here the explicit expressions for $\hat{T}_{\mu\nu}$ and $\hat{N}_\mu$
 and switching to the imaginary-time (Matsubara) formalism via the substitutions $t\to -i\tau$, $\partial_t\to i\partial_\tau$, we obtain
\begin{equation}
\begin{array}{lll}\label{eq:corpm1}
-G^{M}_\zeta(\omega_n) &=& \frac{1}{3}\Pi
[i\gamma_1\partial_1,i\gamma_1\partial_1]+
\frac{2}{3}\Pi
[i\gamma_1\partial_1,i\gamma_2\partial_2]
-2\gamma\Pi
[i\gamma_1\partial_1,-\gamma_0\partial_\tau]\\[1ex]
&-&2\delta\Pi
[i\gamma_1\partial_1,\gamma_0]+2\gamma\delta 
\Pi
[-\gamma_0\partial_\tau,\gamma_0]+
\gamma^2\Pi
[-\gamma_0\partial_\tau,-\gamma_0\partial_\tau]\\[1ex]
&+&\delta^2\Pi[\gamma_0,\gamma_0]+
2(1+\gamma)\Pi
[i\gamma_1\partial_1,i\slashed\partial_\tau-m_0]-
2\gamma(1+\gamma)\Pi
[-\gamma_0\partial_\tau,i\slashed\partial_\tau-m_0]\\[1ex]
&-&2\delta(1+\gamma)\Pi
[\gamma_0,i\slashed\partial_\tau-m_0]+
(1+\gamma)^2\Pi
[i\slashed\partial_\tau-m_0,
i\slashed\partial_\tau-m_0],\quad
\end{array}
\end{equation}
with two-point correlation functions defined as
\bea\label{eq:rings_ab}
\Pi[\hat{a},\hat{b}](\omega_n) =
\int_{0}^{\beta}\!\! d\tau e^{i\omega_n\tau}
\int d\bm r
\langle {\cal T}_\tau(\bar\psi\hat{a}\psi\Big\vert_{(\bm r,\tau)},
\bar\psi\hat{b}\psi\Big\vert_0)\rangle_0,
\eea 
where $\omega_n=2\pi nT$, $n=0,\pm 1,\ldots$ are the bosonic Matsubara
frequencies; ${\cal T}_\tau$ is the imaginary time-ordering operator;
$i\slashed\partial_\tau\equiv
-\gamma_0\partial_\tau-i\gamma_j\partial_j $,
and $\hat{a}$ and $\hat{b}$ are either constants or $\gamma$-matrices
contracted with partial derivatives.  (Note that the correlators which
arise from the interaction part of Equation~\eqref{eq:lagrangian}
vanish for $\omega\neq 0$ because of the energy conservation, see
Ref.~\cite{Harutyunyan:2017b} for details).  Figure~\ref{fig:loops_ab}
illustrates diagrammatically the series of the loop diagrams which
contribute to the correlation function given by
Equation~\eqref{eq:rings_ab}.

\subsection*{Resummation of the Feynman Diagrams}

The class of leading-order diagrams which contributes to the
correlation function~\eqref{eq:rings_ab} is identified according to
the ${\cal O}(1/N_c)$ power-counting scheme. In this scheme each
diagram is selected according to its power with respect to the color
number $N_c$, which is determined by the following
rules~\cite{Quack:1994}: (a)~each quark loop contributes a factor of
$N_c$, which arises from the trace over the color space; (b)~each
coupling $G$ contributes a factor of $1/N_c$. It is easy to see that
the leading-order diagrams in the correlation
function~\eqref{eq:rings_ab} are of the order of ${\cal O}(N_c^1)$ and
involve loop diagrams without vertex corrections, \ie, those of the
type shown in the first and the second lines in
Figure~\ref{fig:loops_ab}. Indeed, \mbox{the factor} $N_c$ associated with each
additional loop is compensated by the factor $1/N_c$ from an interaction
insertion. Therefore, we conclude that we need to resume an infinite
chain of loop diagrams without vertex corrections. \mbox{To carry out} the
resummation, define the single-loop diagram in the momentum space as
\bea\label{eq:ring1_ab}
\Pi_0[\hat{a},\hat{b}](\omega_n)=-
T\sum_l\!\! \int\!\! \frac{d\vecp}{(2\pi)^3}
\Tr \left[\hat{a}D(\vecp, i\omega_l+i\omega_n) 
\hat{b} D(\vecp, i\omega_l) \right],
\eea
where $D(\vecp, i\omega_l)$ is the full (\ie, dressed) quark
propagator defined in the imaginary time, and the summation is over
the fermionic Matsubara frequencies $\omega_l=\pi(2l+1)T-i\mu$,
$l=0,\pm1,\ldots,$ where $\mu$ is the quark chemical potential. The
traces should be taken in Dirac, color, and flavor spaces. \mbox{The
resummation} then leads to
\bea\label{eq:cor_11_ab}
\Pi[\hat{a},\hat{b}]=
\Pi_0[\hat{a},\hat{b}]+\tilde{G}
\Pi_0[\hat{a},\Gamma]\Pi_0[\Gamma,\hat{b}],
\eea
where $\Gamma=1$ (the correlators with the pseudoscalar vertex
$\propto\gamma_5$ vanish because the relevant traces vanish in the
Dirac space). The frequency arguments in Equation~\eqref{eq:cor_11_ab} were
omitted for the sake of brevity. The effective coupling in Equation~\eqref{eq:cor_11_ab}
is related to the bare coupling $G$ via 
\bea\label{eq:G_wave}
\tilde{G}(\omega_n)=\frac{G}{1-G\Pi_0[1,1](\omega_n)}.
\eea
The diagrams involving vertex corrections, such as the one shown in
the third line of Figure~\ref{fig:loops_ab}, are of higher order in the
$O(1/N_c)$ power-counting scheme.  Thus, the computation of the
leading-order contribution to the bulk viscosity reduces to the
calculation of the series of loop diagrams defined by
Equation~\eqref{eq:cor_11_ab}, which in turn requires the evaluation of the
single-loop diagram given by Equation~\eqref{eq:ring1_ab}.
To carry out the sum over the Matsubara frequencies in
Equation~\eqref{eq:ring1_ab} one needs the frequency dependence of the
operators $\hat{a}$ and $\hat{b}$ which arises when
$\hat{a},\hat{b}\propto \partial_\tau$ [see
Equation~\eqref{eq:corpm1}]. Indeed, in the frequency space, such dependence
translates as
$\partial_\tau\to -i\bar{\omega}_l \equiv -i(\omega_l+\omega_n/2)$.
For such cases, we separate the $i\bar\omega_l$-dependence by formally
factorizing the frequency dependence into a function
$f(i\bar\omega_l)$, \ie, we write
$\hat{a}...\hat{b}...= f(i\bar\omega_l)\hat{a}_0...\hat{b_0}...$,
where $\hat{a}_0$ and $\hat{b}_0$ do not depend on $i\bar\omega_l$.
After summation over the Matsubara frequencies and subsequent
analytical continuation (\ie, $i\omega_n = \omega+i\delta$) we obtain (\mbox{see
Appendix} \ref{app:sums} for details)
\bea\label{eq:reloop}
\Pi_0[\hat{a},\hat{b}](\omega)=
\int_{-\infty}^{\infty}\!\! d\varepsilon 
\int_{-\infty}^{\infty}\!\! d\varepsilon'
\int\frac{d\bm p}{(2\pi)^3}
\Tr[\hat{a}_0 A(\bm p, \varepsilon')
\hat{b}_0  A(\bm p, \varepsilon)]
\frac{\tilde{n}(\varepsilon')f(\varepsilon'-\omega/2) 
-\tilde{n}(\varepsilon)f(\varepsilon+\omega/2)}
{\varepsilon-\varepsilon' +\omega+i\delta},
\eea 
where ${\tilde n}(\varepsilon)=n(\varepsilon)-1/2$ with
$n(\varepsilon)=[e^{\beta(\varepsilon-\mu)}+1]^{-1}$ being the Fermi
distribution for quarks.  Finally, we separate the real and imaginary
parts in Equation~\eqref{eq:reloop} by exploiting the Dirac identity
\be\label{eq:dirac}
\frac{1}{x+i\delta}=P\frac{1}{x}-i\pi\delta(x).
\ee
From Equations~\eqref{eq:reloop} and \eqref{eq:dirac} we find
\bea\label{eq:difimsums}
\frac{d}{d\omega}{\rm Im} \Pi_0[\hat{a},\hat{b}](\omega)
\bigg|_{\omega=0}\!\! &=& -
\pi\int_{-\infty}^{\infty}\!\! d\varepsilon
\int\!\! \frac{d\vecp}{(2\pi)^3}
\frac{\partial n(\varepsilon)}{\partial \varepsilon } f(\varepsilon)
\Tr[\hat{a}_0 A(\bm p, \varepsilon)
\hat{b}_0 A(\bm p, \varepsilon)],
\eea
and
\bea\label{eq:difresums_ab}
{\rm Im} \Pi_0[\hat{a},\hat{b}](\omega)
\bigg|_{\omega=0} =
\frac{d}{d\omega}{\rm Re} \Pi_0[\hat{a},\hat{b}]
(\omega)\bigg|_{\omega=0}=0.
\eea
Using Equation~\eqref{eq:difresums_ab} we can compute
the imaginary part of Equation~\eqref{eq:cor_11_ab}
\bea\label{eq:difcor_ab}
\frac{d}{d\omega}{\rm Im}\Pi
[\hat{a},\hat{b}](\omega)\bigg\vert_{\omega=0}
=L_0[\hat{a},\hat{b}] + \bar GL_1[\hat{a},\hat{b}]+\bar G^2L_2[\hat{a},\hat{b}],
\eea
where 
\bea
L_0[\hat{a},\hat{b}] &=&
\frac{d}{d\omega}{\rm Im}\Pi_0
[\hat{a},\hat{b}](\omega)\Big\vert_{\omega=0},\\
L_1[\hat{a},\hat{b}] &=& 
R_0[\hat{a},1]L_0[1,\hat{b}]
+R_0[1,\hat{b}]L_0[\hat{a},1],\quad\\
L_2[\hat{a},\hat{b}] &=&  L_0[1,1]R_0[\hat{a},1]
R_0[1,\hat{b}],\\
R_0[\hat{a},\hat{b}] &=& {\rm Re}\Pi_0 [\hat{a},\hat{b}](\omega)\Big\vert_{\omega=0},\\
\label{eq:G_bar}
\bar{G}
&=& \frac{G}{1-GR_0[1,1]}.
\eea

To compute the traces in Equations~\eqref{eq:reloop} and \eqref{eq:difimsums}
one needs to exploit the Dirac decomposition of the spectral function
\bea\label{eq:spectralfunction}
A(\bm p,p_0)
=-\frac{1}{\pi}(mA_s+p_0\gamma_0 A_0 
-\bm p\bm\gamma A_v),
\eea 
where $m$ is the quark mass.  The coefficients $A_s$, $A_0$ and $A_v$
can be expressed in terms of the relevant components of the quark
self-energy according to the
relations~\cite{Lang:2015,LW14,Harutyunyan:2017a}
\bea\label{eq:spectral_coeff}
A_i(p_0, p)=\frac{1}{d} 
[n_1\varrho_i -2n_2 (1+r_i) ],\qquad 
d=n_1^2+4n_2^2,
\eea
where $\varrho_i = {\rm Im}\Sigma_i$,
$r_i = {\rm Re}\Sigma_i$, $i=s,0,v$, and
\bea
\label{eq:N1}
n_1&=&p_0^2[(1+r_0)^2-\varrho_0^2]
-\bm p^2[(1+r_v)^2-\varrho_v^2]
-m^2[(1+r_s)^2-\varrho_s^2],\\
\label{eq:N2}
n_2&=&p_0^2\varrho_0 (1+r_0) 
-\bm p^2\varrho_v(1+r_v)-m^2\varrho_s(1+r_s).
\eea

From now on we will neglect the irrelevant real parts of the
self-energy, which lead to momentum-dependent corrections to the
constituent quark mass in next-to-leading order ${\cal O} (N_c^{-1})$
and will keep only the imaginary parts which were computed in
Refs.~\cite{Lang:2015,LW14,Harutyunyan:2017a}.  The three components of the
quark self-energy are identified via
\bea\label{eq:selfenergy}
\Sigma^{R(A)}
=m\Sigma_s^{(*)}
-p_0\gamma_0\Sigma_0^{(*)}
+\bm p\bm\gamma\Sigma_v^{(*)}.
\eea 

With this input, we can now calculate the relevant correlators entering
Equation~\eqref{eq:corpm1}. Computing the traces and performing 
the angular integrations in Equations~\eqref{eq:reloop} and \eqref{eq:difimsums},
we find, \eg,
\bea\label{eq:reloop_generic}
R_0\left[1,i\gamma_1\partial_1 \right]
&=&-\frac{2N_cN_f}{3\pi^4}\!\!
\int_{-\infty}^{\infty}\!\!\! d\varepsilon
\int_{-\infty}^{\infty}\!\!\! d\varepsilon'
\int_0^\Lambda\!\!\! dp
\frac{{n}(\varepsilon)-{n}(\varepsilon')}
{\varepsilon-\varepsilon'}mp^4(A_s'A_v+A_sA_v'),\\
\label{eq:difimloop_final}
L_0[i\gamma_1\partial_1,i\gamma_1\partial_1] &=&
 -\frac{2N_cN_f}{15\pi^3}\int_{-\infty}^{\infty} 
d\varepsilon\int_0^\Lambda dp\frac{\partial n(\varepsilon)}{\partial\varepsilon}
p^4(-5m^2A_s^2+5\varepsilon^2 A_0^2+p^2A_v^2),
\eea
where $A'_i\equiv A_i(\bm p, \varepsilon')$; $N_c=3$ and $N_f=2$ are
the color and flavor numbers, respectively.  The remaining correlation
functions can be computed in analogy to Equations~\eqref{eq:reloop_generic}
and \eqref{eq:difimloop_final}; the explicit expressions are given in
Ref.~\cite{Harutyunyan:2017b}.
 
Inserting the relevant correlators in Equations~\eqref{eq:bulk_kubo}, \eqref{eq:corpm1}, \eqref{eq:difcor_ab}--\eqref{eq:G_bar}, we can write the bulk \mbox{viscosity as} 
\bea\label{eq:zeta_sum}
\zeta = \zeta_0 +\zeta_1 +\zeta_2, 
\eea 
where each of the three terms arises from the corresponding terms in
Equation~\eqref{eq:difcor_ab}. The first (single-loop) term is given
by
\bea\label{eq:zeta0_final} 
\zeta_0 =
-\frac{2N_cN_f}{9\pi^3} \int_{-\infty}^{\infty} d\varepsilon
\frac{\partial n}{\partial\varepsilon} \int_0^\Lambda dpp^2
 \Big[2(ax+by+cz)^2 -(x^2-y^2+z^2)(a^2-b^2+c^2)\Big],
\eea
where 
$x=3(1+\gamma)m_0,~ y=3(\delta-\varepsilon), ~ z=(2+3\gamma)p,$
$a=mA_s,~ b=\varepsilon A_0,$ and $ c=pA_v.$
 The multiloop contributions are given by
\bea\label{eq:zeta_12_final}
\zeta_1 =2(\bar G\bar R )I_1,\qquad \zeta_2 =  (\bar G\bar R )^2I_2,
\eea
where the effective coupling $\bar G$ is given by
\bea\label{eq:G_bar0}
\bar G = \frac{G}{1-R_0G},
\eea
with the polarization loop
\bea\label{eq:R_0}
R_0 = -\frac{2N_cN_f}{\pi^4}
\int_{-\infty}^{\infty} d\varepsilon 
\int_{-\infty}^{\infty} d\varepsilon' 
\frac{n(\varepsilon)-n(\varepsilon')}
{\varepsilon-\varepsilon'}
 \int_0^\Lambda dpp^2 (aa'+bb'-cc'),
\eea
and
\bea \label{eq:I1}
I_1&=&-\frac{2N_cN_f}{3\pi^3}
\int_{-\infty}^{\infty} d\varepsilon 
\frac{\partial n}{\partial\varepsilon}
\int_0^\Lambda dpp^2
\Big[x(a^2+b^2-c^2)+2a(by+cz)\Big],\\
\label{eq:I2}
I_2&=&-\frac{2N_cN_f}
{\pi^3 }\int_{-\infty}^{\infty} 
d\varepsilon \frac{\partial n}{\partial\varepsilon} \int_0^\Lambda dpp^2 
(a^2+b^2-c^2),\\
\label{eq:R_bar}
\bar R &=&-\frac{2N_cN_f}{3\pi^4}
\int_{-\infty}^{\infty} d\varepsilon 
\int_{-\infty}^{\infty} d\varepsilon' 
\int_0^\Lambda dp {p^2}
\frac{1}{\varepsilon-\varepsilon'}
\Big\{[n(\varepsilon)-n(\varepsilon')]\nonumber\\
&\times&
\big[x(aa'+ bb'-cc')+z(a'c+ac')\big]
+\Big[yn(\varepsilon)-y'n(\varepsilon')+\frac{3}{2}(\varepsilon-\varepsilon')\Big]
(a'b+ ab')\Big\}.
\eea 
Here the functions $a',b',c',y'$ are obtained from $a,b,c,y$ defined
above by substitution $\varepsilon\to\varepsilon'$.  Equations
\eqref{eq:zeta_sum}--\eqref{eq:R_bar} express the bulk viscosity of
 quark plasma in terms of the components of its spectral function.

 It is remarkable that the multiloop contributions to the bulk
 viscosity vanish trivially in the chirally symmetric case, where
 $m_0=0$. Indeed, for massless quarks and temperatures $T>T_c$, where
 $T_c$ is the critical temperature of the chiral phase transition, we
 find $x=0$ and $a, a'\propto m=0$ in Equation~\eqref{eq:R_0}.
 \mbox{As a result}, we have $\zeta_{1,2}=0$ and $\zeta=\zeta_0$
 according to Equations~\eqref{eq:zeta_12_final} and
 \eqref{eq:zeta_sum}.
 
\section{Numerical Results}
\label{sec:results}

We use the Lorentz components of the quark spectral function, obtained
previously in Refs.~\cite{Lang:2015,LW14,Harutyunyan:2017a}, to evaluate
numerically the bulk viscosity. We concentrate on the region of the
phase diagram which is located above the Mott transition temperature
$T_{\rm M}$, which is defined as the threshold temperature at any
given chemical potential above which the meson decay into two
on-mass-shell quarks is kinematically allowed. It is identical to the
chiral phase transition temperature $T_c$ in the chiral limit $m_0=0$.

\begin{figure}[H] 
\begin{center}
\includegraphics[height=6.cm,keepaspectratio]{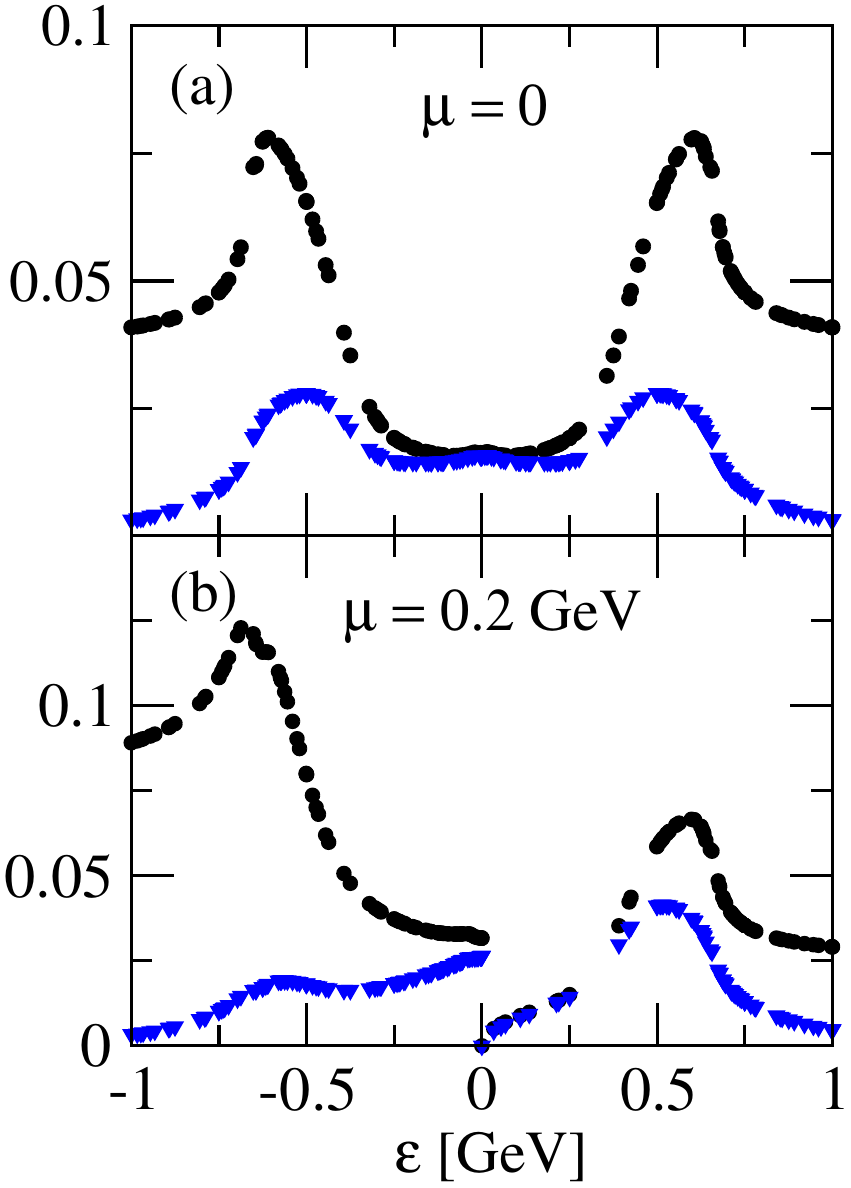}
\hskip 0.3cm
\includegraphics[height=6.cm,keepaspectratio]{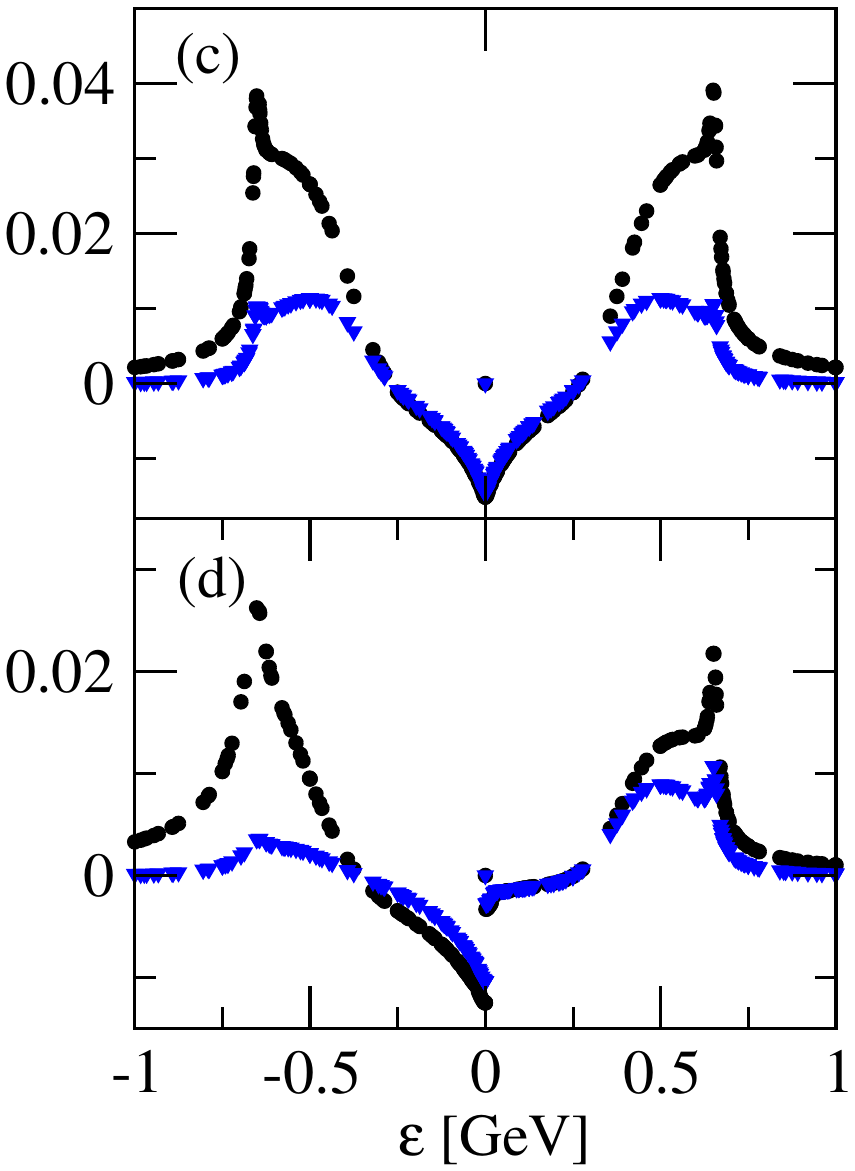}
\hskip 0.3cm
\includegraphics[height=6.cm,keepaspectratio]{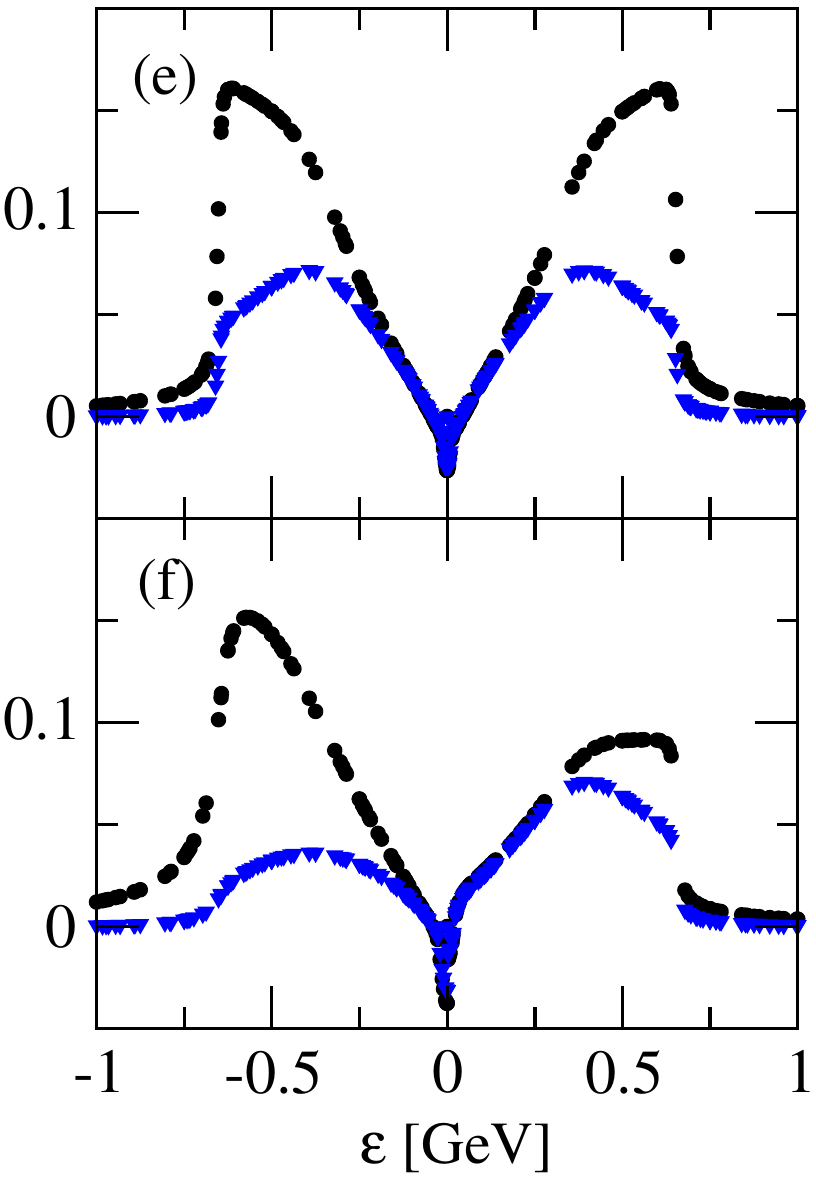}
\vspace{-6pt}
\caption{ The integrands of $\zeta_0$ (\textbf{a},\textbf{b}), $I_1$ (\textbf{c},\textbf{d}) and $I_2$ (\textbf{e},\textbf{f}) as functions of the quark energy
  without (\mbox{black circles}, in GeV units) \ and with (blue triangles)
  the factor $-\partial n/\partial\varepsilon$ for vanishing (\textbf{a},\textbf{c},\textbf{e}) and finite (\textbf{b},\textbf{d},\textbf{f}) chemical potential. The
  temperature is fixed at $T=0.26$ GeV.  }
\label{fig:itgr1_zeta0} 
\end{center}
\end{figure}

To gain insight into the numerical results for the bulk viscosity, it
is useful first to analyze the integrands of
Equations~\eqref{eq:zeta0_final} and
\eqref{eq:R_0}--\eqref{eq:R_bar}. The integrands of $\zeta_0$, $I_1$
and $I_2$ are shown in Figure~\ref{fig:itgr1_zeta0}.  Each of these
integrands develops a peak structure at $p\simeq |\varepsilon|$,
whereby the height of the peak rapidly increases with
$|\varepsilon|$. \mbox{The momentum} integrals are increasing
functions of $|\varepsilon|$ as long as
$\vert\varepsilon\vert\leq\Lambda$, and decreasing for larger energies
(because of the momentum cut-off $p\leq\Lambda$, see
Figure~\ref{fig:itgr1_zeta0}).

Note that for $\mu= 0$ the integrands
are even functions of $\varepsilon$, which reflects the
quark-antiquark symmetry.  The factor
$\partial n(\varepsilon)/\partial\varepsilon$ breaks this symmetry for
non-vanishing chemical potentials by increasing the contribution of
quarks. Thus, we see that the dominant contribution to the bulk viscosity
comes from the modes with $p\simeq |\varepsilon|$, whereby the quark
contribution dominates the antiquark contribution at non-zero $\mu$.

\begin{figure}[H] 
\begin{center}
\includegraphics[width=6.5cm,keepaspectratio]{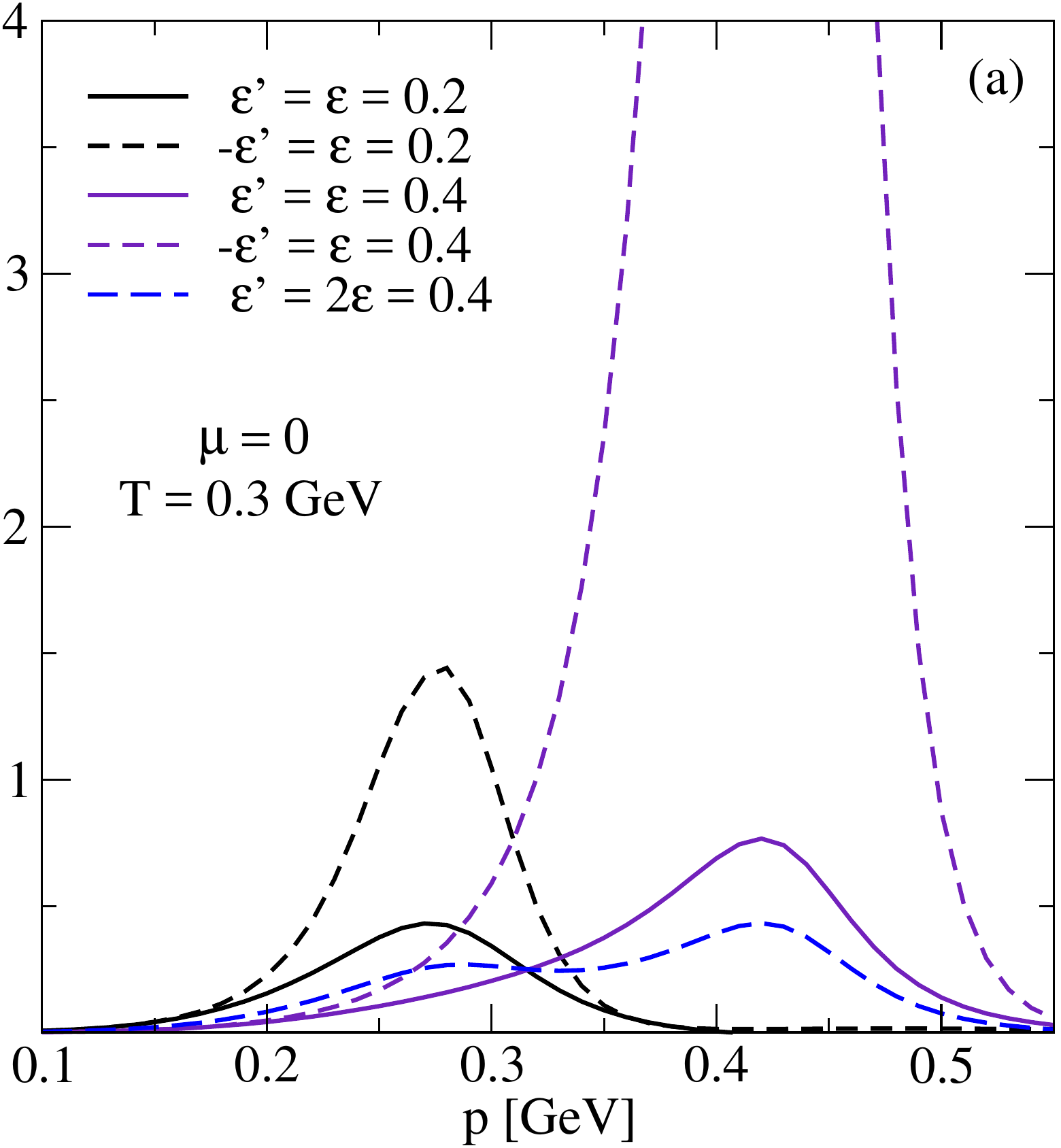}
\hspace{1.5cm}
\includegraphics[width=6.57cm,keepaspectratio]{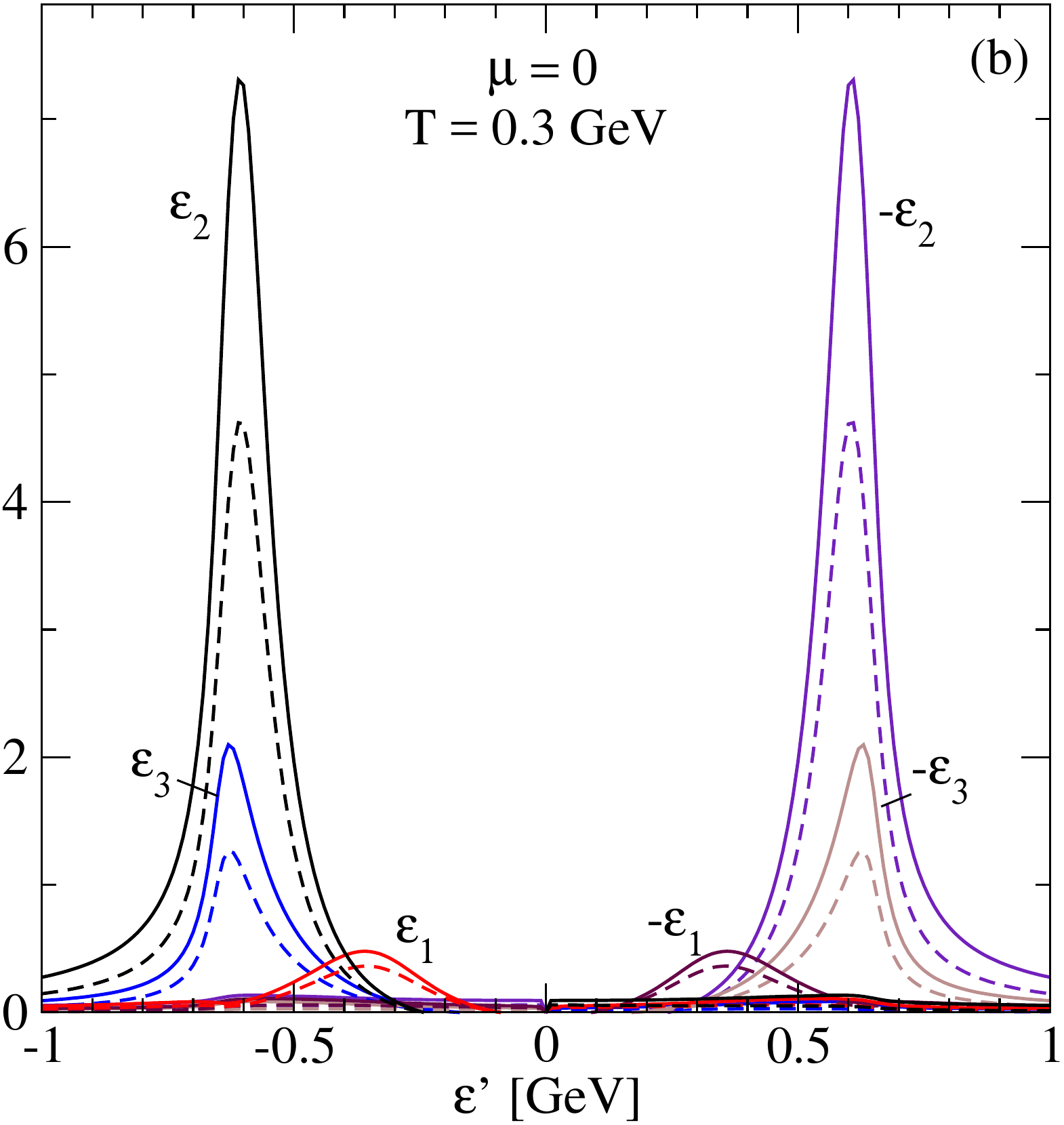}
\vspace{-6pt}
\caption{ The integrands of the integral $R_0$: (\textbf{a}) the inner
  integrand as a function of quark momentum at various values of
  $\varepsilon$ and $\varepsilon'$ (shown in GeV units); (\textbf{b}) the
  $p$-integral of $R_0$ (solid lines, in GeV units) and its product
  with the factor
  $[n(\varepsilon)-n(\varepsilon')]/(\varepsilon-\varepsilon')$
  (dashed lines) as functions of $\varepsilon'$ for various values of
  $\varepsilon$: $\varepsilon_1=0.3$, $\varepsilon_2=0.6$ and
  $\varepsilon_3=0.7$ GeV.}
\label{fig:itgr_R0} 
\end{center}
\end{figure}

Now we turn to the three-dimensional integrals $R_0$ and $\bar R$
given by Equations~\eqref{eq:R_0} and \eqref{eq:R_bar}. Their integrands
are strongly peaked at $p\simeq |\varepsilon|\simeq|\varepsilon'|$,
and have two smaller maxima located at $p\simeq |\varepsilon|$ and
$p\simeq |\varepsilon'|$ in the cases where
$|\varepsilon|\neq|\varepsilon'|$, see Figure~\ref{fig:itgr_R0}.  As a
result, the momentum integrals of these expressions obtain the main
contribution form energies $\varepsilon'\simeq\pm\varepsilon$. We also
observe that the height of each peak increases with $|\varepsilon|$
for $|\varepsilon|\le \Lambda$ and sharply drops beyond the
cut-off. The peak structures seen above reflect the quasiparticle-like
nature of the excitations, which however have non-zero width because
of the meson decay and recombination processes included in our
consideration.

\begin{figure}[H] 
\begin{center}
\includegraphics[width=6.5cm,keepaspectratio]{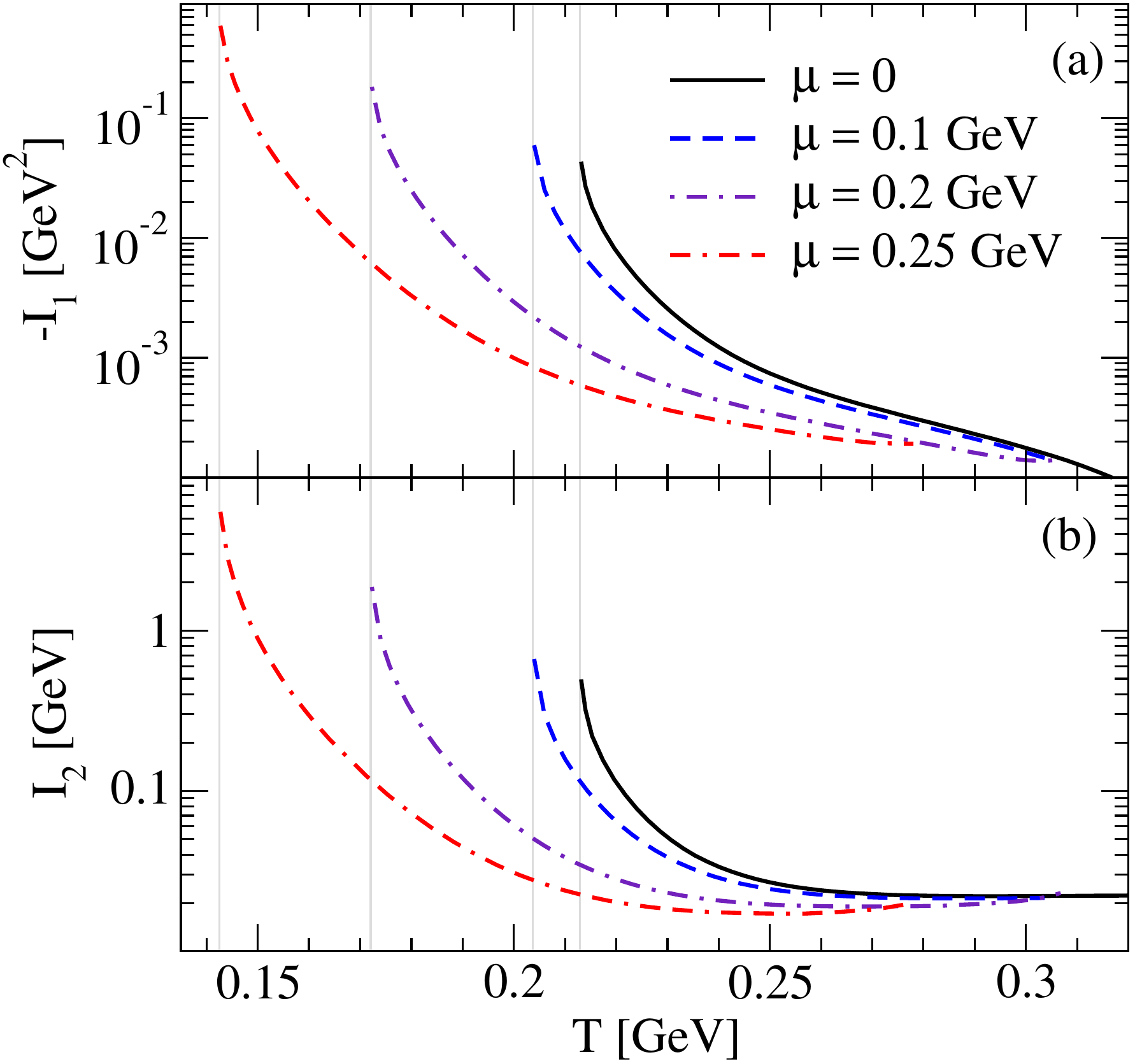}
\includegraphics[width=6.55cm,keepaspectratio]{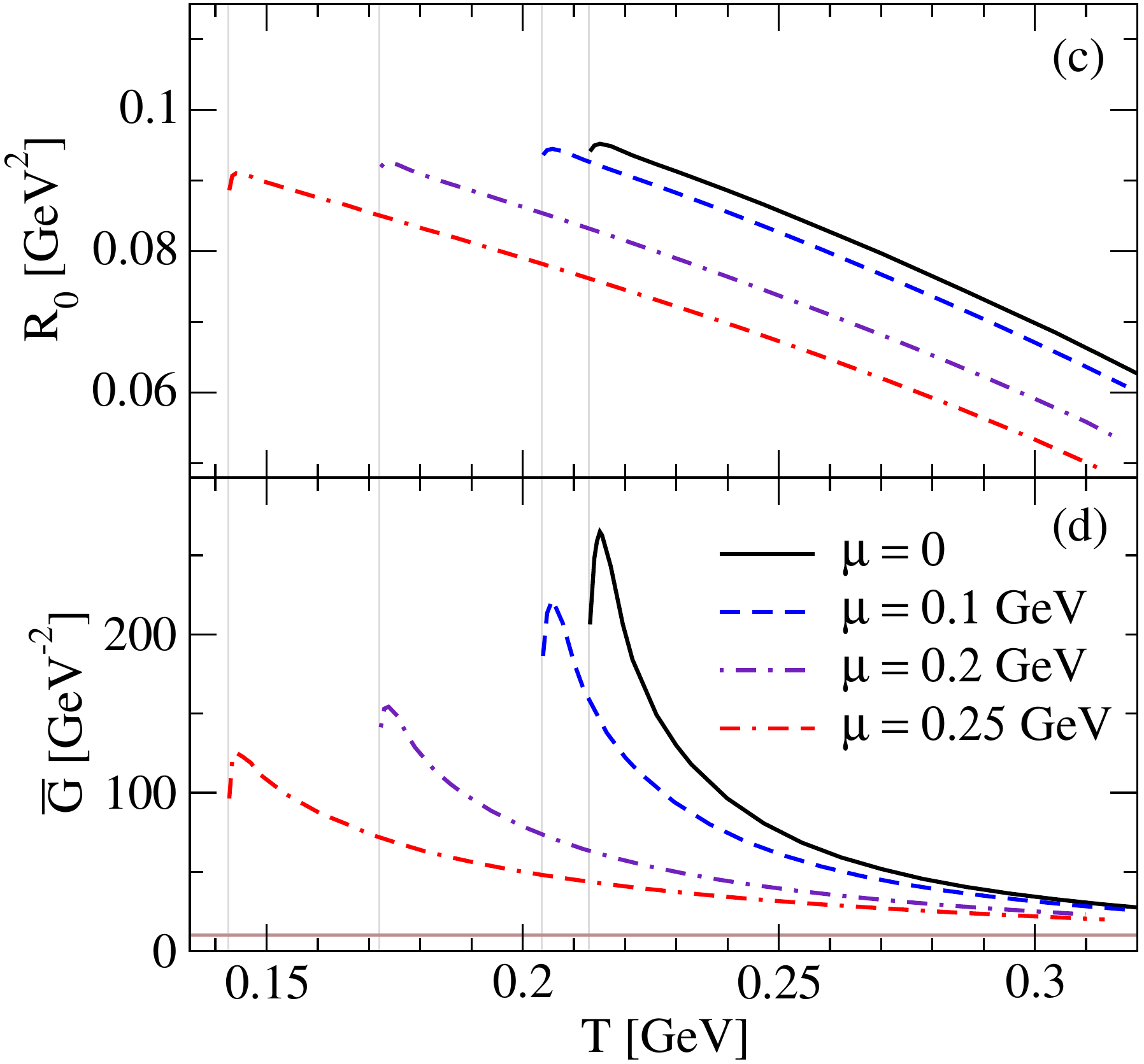}
\vspace{-6pt}
\caption{Dependence of the integrals $I_1$ (\textbf{a}), $I_2$ (\textbf{b}), $R_0$ (\textbf{c})
  and the renormalized coupling $\bar G$ (\textbf{d}) on the temperature for
  several values of the chemical potential. The corresponding Mott
  lines are shown by vertical lines.  The value of the bare coupling
  constant $G$ is shown by the solid horizontal line.  }
\label{fig:I12_R0_G} 
\end{center}
\end{figure}

Figure \ref{fig:I12_R0_G} illustrates the temperature and chemical
potential dependence of the integrals $I_1$, $I_2$, $R_0$ and the renormalized
coupling $\bar G=G/(1-GR_0)$ [given by Equation~\eqref{eq:G_bar}].  The behavior
of $\bar R$ is similar to $R_0$ and is not shown.  Quantitatively, the
three- and two-dimensional integrals $R_0$ and $I_1,I_2$,
respectively, show the same behavior, reflecting the importance of the
meson decay processes close to the Mott line. The renormalized
coupling $\bar G$ attains its maximum close to the Mott line, where it
exceeds the bare coupling constant roughly by an order of
magnitude. Because of this behavior of $\bar{G}$, the multiloop
contributions to the bulk viscosity given by
Equation~\eqref{eq:zeta_12_final} are expected to be important close to the
Mott temperature.

\begin{figure}[H] 
\begin{center}
\includegraphics[width=7cm,keepaspectratio]{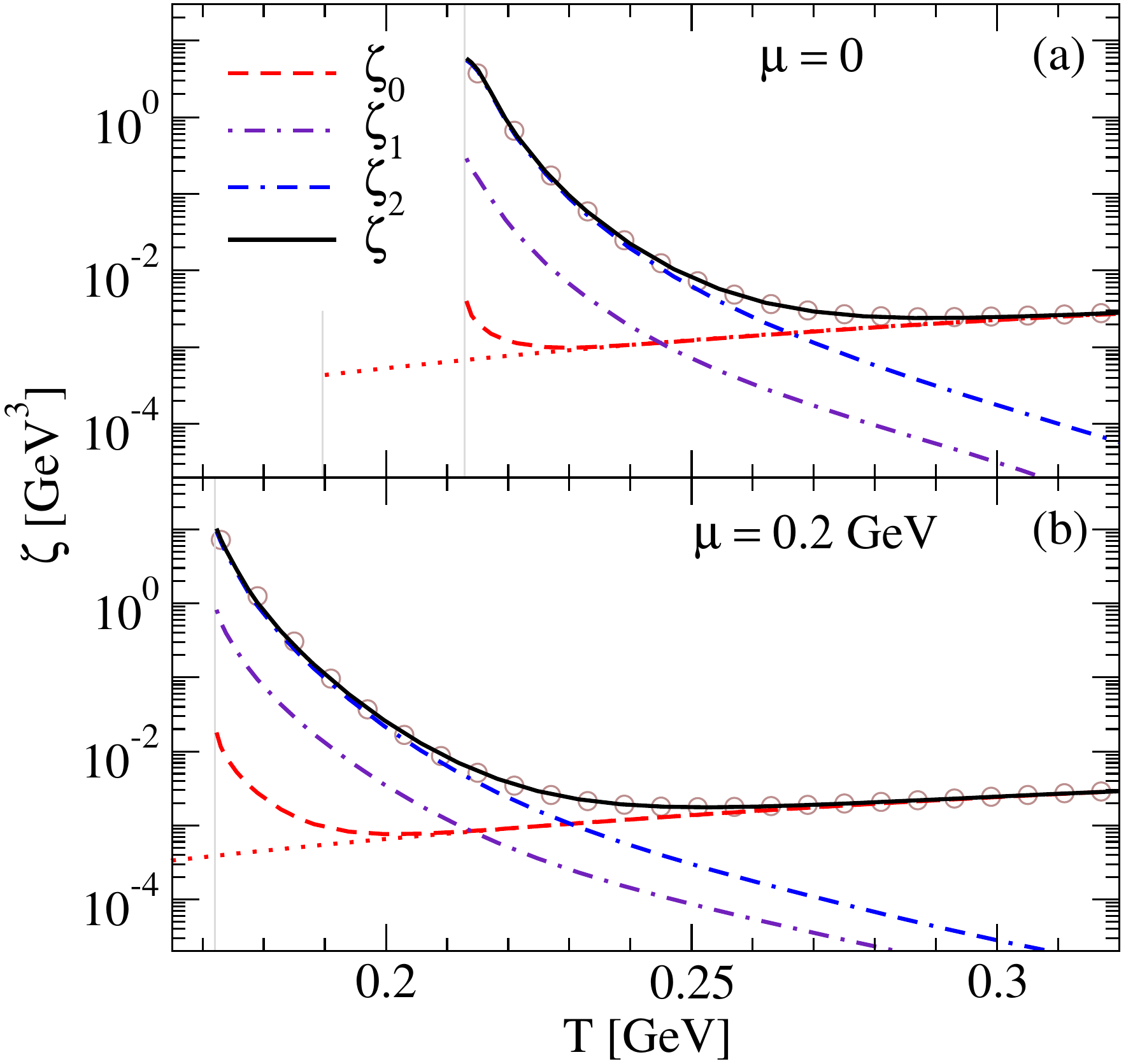}
\vspace{-6pt}
\caption[]{ The contributions of the single-loop ($\zeta_0$) and
  multiloop ($\zeta_1$, $\zeta_2$) diagrams to the bulk viscosity, as
  well as their sum as functions of the temperature for vanishing (\textbf{a})
  and finite (\textbf{b}) chemical potential.  The dotted lines correspond to
  the chiral limit $m_0=0$.  The results of the fit formula
  \eqref{eq:fit_zeta} are \mbox{shown by circles.}}
\label{fig:zeta} 
\end{center}
\end{figure}

The results for the bulk viscosity are shown in
Figure~\ref{fig:zeta}. The multiloop contributions $\zeta_{1}$ and
$\zeta_{2}$ dominate over the one-loop contribution $\zeta_0$ in the
regime close to the Mott transition line, where all three components
rapidly decrease with the temperature and density. At high enough
temperatures, \mbox{the one-loop} contribution scales as $T^3$ and
dominates over the multiloop contributions.  \mbox{The net bulk}
viscosity which is the sum of the one-loop and multiloop contributions
then exhibits a shallow minimum as a function of temperature. From the
analysis above, we thus conclude that the single-loop approximation is
justified only at sufficiently high temperatures where multiloop
diagrams \mbox{are suppressed.}

We now comment briefly on the case where the chiral symmetry is
intact, \ie, $m_0=0$. In this case, the multiloop contributions vanish
automatically, as explained in Section~\ref{sec:Kubo_bulk}. The bulk
viscosity is then given by the single-loop contribution $\zeta_0$
taken in the limit $m\to 0$, which is shown in Figure~\ref{fig:zeta} by
the dotted lines for zero and finite chemical potentials.  Contrary to
the case where $m_0\neq 0$, here $\zeta_0$ is smooth at the Mott
temperature and increases with the temperature $\propto T^3$ in the
whole temperature-density range. We thus conclude that the {\it explicit}
chiral symmetry breaking is essential for the correct description of
the bulk viscosity in the low-temperature region of the phase
diagram, especially in the region close to the chiral phase transition line.

\begin{figure}[H] 
\begin{center}
\includegraphics[width=7cm,keepaspectratio]{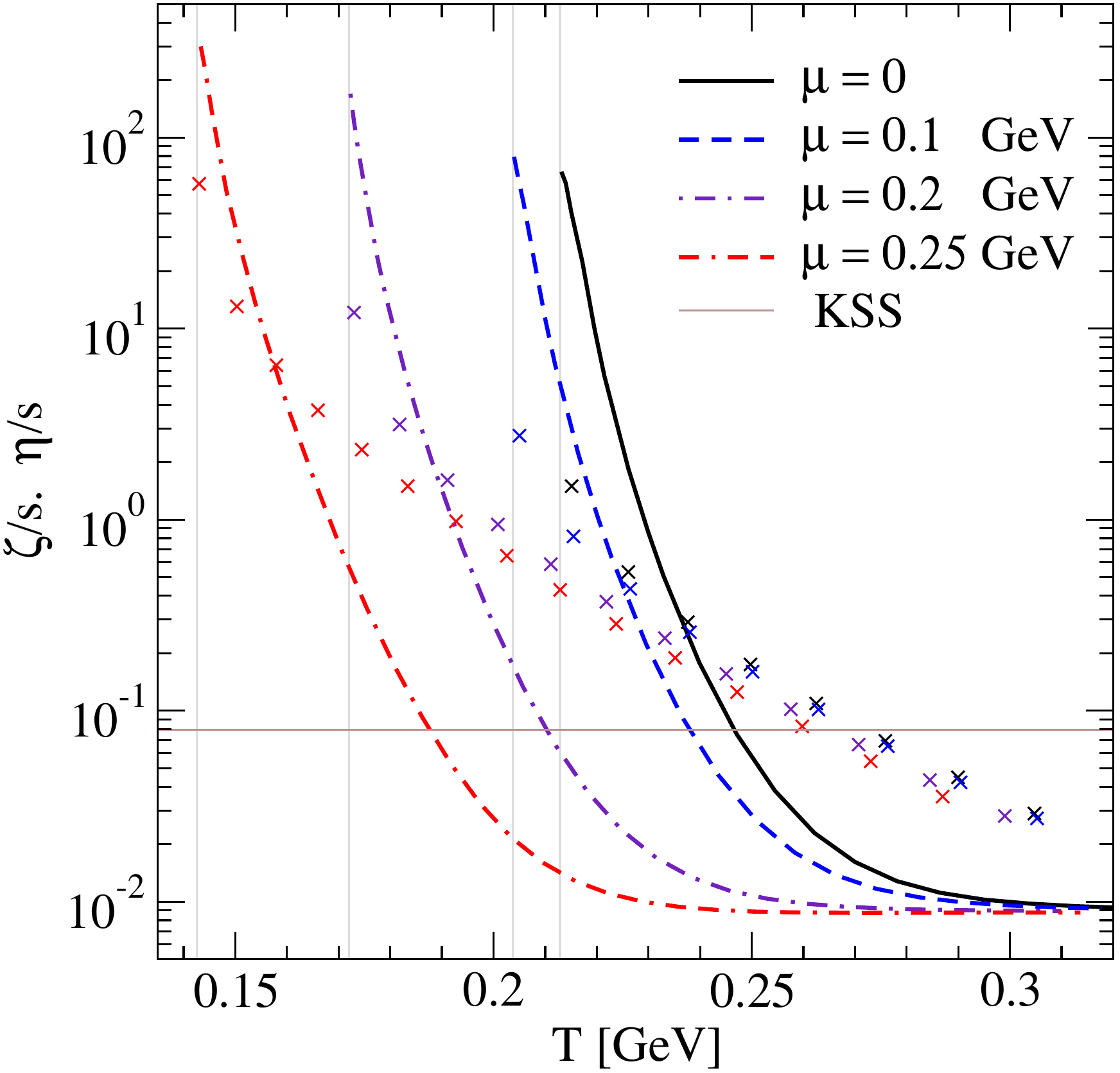}
\vspace{-6pt}
\caption[]{ The ratio $\zeta/s$ as a function of the temperature for
  several values of the chemical potential. The corresponding $\eta/s$
  ratios are shown for comparison by crosses. The solid horizontal
  line shows the KSS bound~\cite{Kovtun:2005}.}
\label{fig:zeta_s} 
\end{center}
\end{figure}

For completeness, we also compare our results with the shear viscosity
$\eta$, which was computed previously in Ref.~\cite{Harutyunyan:2017a}
(see also Refs.~\cite{LW14,Lang:2015}) by employing the same formalism
and approximations.  Figure \ref{fig:zeta_s} shows the dependence of
the ratios $\zeta/s$ and $\eta/s$, where $s$ is the entropy density,
on the temperature for several values of the chemical
potential~\cite{Harutyunyan:2017a,Harutyunyan:2017b}.  For comparison,
we also show the AdS/CFT lower bound $1/4\pi$ on the $\eta/s$
ratio~\cite{Kovtun:2005}.  We see that both ratios decrease rapidly
with the temperature, but the slope of this decrease is larger in
the case of the bulk viscosity in the region which is close to the
Mott transition line.  In this regime, the bulk viscosity exceeds the
shear viscosity by factors $\zeta/\eta \simeq 5 \div 20$. Thus, in the
low-temperature regime close to the Mott transition line the bulk
viscosity is the dominant source of dissipation.  It is worth
stressing that had we kept only the one-loop contribution to the bulk
viscosity, it would have been negligible compared to the shear
viscosity.  As $\zeta$ drops much faster than $\eta$ with the
temperature, the shear viscosity is the dominant dissipation channel
at high temperatures.

Our numerical results can be fitted using the formula
\bea\label{eq:fit_zeta}
\zeta_{\rm fit}(T,\mu)=a(y)\exp\left[\frac{c(y)}{T/T_M(y)-b(y)}\right]+d(y)T^3,
\eea
where $y=\mu/\mu_0$, and $\mu_0=0.345$ GeV is the  value of the  chemical potential at which the Mott line terminates. The coefficients $a,b,c,d$ depend only on the chemical potential and are given by
\bea\label{eq:coeff_abc}
a(y) &=& (2.57 - 5.65 y^2)\times 10^{-6}\hspace{0.1cm} [{\rm GeV}^3],\\
b(y) &=& 0.806 - 0.055 y^2 - 0.617 y^4,\\
c(y) &=& 2.89 + 0.96 y^2 +12.73 y^4,\\
d(y) &=& 0.082 + 0.02 y^2.
\eea

The relative error of the fit formula \eqref{eq:fit_zeta} is
$\le 10\%$ for chemical potentials $\mu\le 0.2$ GeV. The bulk
viscosity according to our fit formula is shown in Figure~\ref{fig:zeta}
by empty circles.  The fit formula above should be complemented by a
fit to the Mott temperature as a function of the chemical potential,
which is given by the formula
\bea\label{eq:fit_mott}
\hspace{-0.2cm}
T_{\rm M}^{\rm fit}(\mu)=T_0
\left\{\begin{array}{ll} 1-\sqrt{\gamma y}
e^{-\pi/(\gamma y)}         &0\leq y\leq 0.5,\\
         \sqrt{1.55(1-y)+0.04(1-y)^2}  &0.5< y\leq
                                          1,\end{array}\right.
\eea
where $T_0=T_{\rm M}(\mu=0)=0.213$ GeV, and $\gamma =2.7$.  
The relative accuracy of the formula \eqref{eq:fit_mott} is $\le 3\%$ 
for chemical potentials $\mu\le 0.32$ GeV.

\section{Conclusions}
\label{sec:conclusions}

In this contribution, we provided a general derivation of the Kubo
formula for the bulk viscosity from Zubarev's formalism of NESO
generalized to systems with multiple conserved charges. \mbox{The method} was
then illustrated on the example of computation of the bulk viscosity
of quark matter in the framework of the two-flavor NJL model. The
previous discussion of Ref.~\cite{Harutyunyan:2017b} has been
supplemented by further details.

The key finding of our work is that at low temperatures and close to
the Mott transition line the overall multiloop contribution to the
bulk viscosity is larger than the one-loop contribution.  This is in
contrast to the results found for the shear viscosity and the thermal
and electrical conductivities, for which the single-loop approximation
gives the leading-order result~\cite{Harutyunyan:2017a}. We have shown
that the bulk viscosity decreases with the temperature and the
chemical potential in this regime, attains a minimum and then
increases again at higher temperatures where the one-loop contribution
becomes dominant.

\textls[-15]{Phenomenologically interesting is the fact that the bulk viscosity
provides the main source of dissipation of stresses close to the Mott
line as it exceeds the shear viscosity in this regime by factors of
$5\div 20$. The bulk viscosity drops faster than the shear viscosity
as the temperature increases and it becomes negligible above a certain
value of the temperature.  Finally, we observed that in the chiral
symmetric case, where the quark masses vanish above the (critical)
Mott temperature, the picture is different. In this case, the
multiloop contributions to the bulk viscosity vanish, and
consequently the bulk viscosity becomes negligible compared to the
shear viscosity in the entire \mbox{temperature-density plane.}}

\vspace{6pt}

\authorcontributions{\textls[-20]{Conceptualization, A.H., A.S.;
Methodology, A.H., A.S.;
Investigation, A.H., A.S.;
Writing—Original Draft Preparation, A.H., A.S.;
Writing—Review \& Editing, A.H., A.S.;
Funding Acquisition, A.S.}}
\funding{This research was funded by 
Deutsche Forschungsgemeinschaft, grant number SE~1836/4-1 and by European
Cooperation in Science and Technology (Actions MP1304 and CA16214).}

\acknowledgments{We are grateful to Xu-Guang Huang and Dirk H. Rischke
  for discussions.  ~A.H. acknowledges support from the HGS-HIRe
  graduate program at Goethe University.~A.S.~was supported by the
  Deutsche~Forschungsgemeinschaft (Grant No.~SE~1836/4-1).  The
  support from European COST Actions ``NewCompStar'' (MP1304) and
  "PHAROS" (CA16214) and the LOEWE-Program of Helmholtz International
  Center for FAIR of the state of Hesse (Germany) is gratefully
  acknowledged.  }
  
  \conflictsofinterest{The authors declare no conflict of interest.}

\appendixtitles{yes} 
\appendixsections{multiple} 
\appendix
\section{Matsubara Summations}
\label{app:sums}

\begin{figure}[H]  
\begin{center}
\vskip 0.5cm
\includegraphics[width=8cm,keepaspectratio]{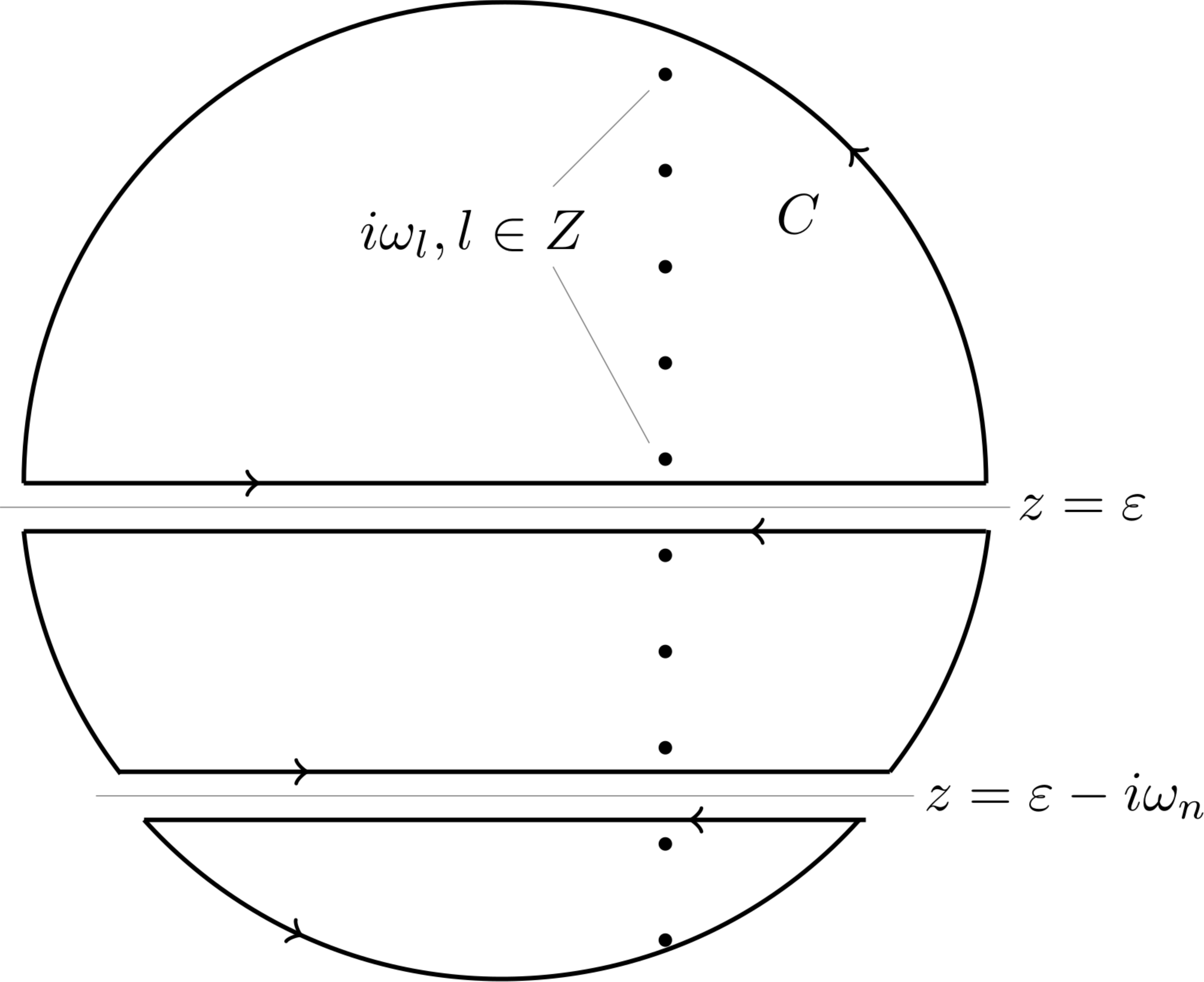}
\vspace{-6pt}
\caption{ The contour of integration in Equation~\eqref{eq:sums_ab}. The dots correspond to the fermionic Matsubara frequencies.}
\label{fig:circle}
\end{center}
\end{figure}

To perform the Matsubara summation in Equation~\eqref{eq:ring1_ab} 
we express the full quark propagator in terms of the quark spectral function
\be\label{eq:propagator}
D(\bm p, z)=\int_{-\infty}^{\infty}d\varepsilon 
\frac{A(\bm p, \varepsilon)}{z-\varepsilon},
\ee 
where the spectral function is defined as  
\be\label{eq:spectralfunction0}
A(\bm p, \varepsilon)=-\frac{1}{2\pi i}\left[D^R(\bm p, \varepsilon)-
D^A(\bm p, \varepsilon)\right],
\ee
with $D^{R/A}(\bm p, \varepsilon)$ being the retarded/advanced Green's functions. According
to Equation~\eqref{eq:propagator}, $D(\bm p, z)$ has a branch cut on the 
real axis, therefore a calculation of the residues gives for 
the integrand of Equation~\eqref{eq:ring1_ab}
\bea\label{eq:sums_ab}
S[\hat{a},\hat{b}](\bm p, i\omega_n)&\equiv & 
T\sum\limits_l 
\Tr\bigl[\hat{a} D(\bm p, i\omega_l+i\omega_n)
\hat{b} D(\bm p, i\omega_l)\bigr]\nonumber\\
&=& T\sum\limits_l f(i\bar\omega_l)
\Tr\bigl[\hat{a}_0 D(\bm p, i\omega_l+i\omega_n)
\hat{b}_0 D(\bm p, i\omega_l)\bigr]\nonumber\\
&=& -\int_C\frac{dz}{2\pi i}{\tilde n}(z)f(z+i\omega_n/2)
\Tr\bigl[\hat{a}_0  D(\bm p, z+i\omega_n)
\hat{b}_0  D(\bm p, z)\bigr],
\eea
where $n(z)=[e^{\beta (z-\mu)}+1]^{-1}$ is 
the Fermi distribution function, and ${\tilde n}(z)=n(z)-1/2$. The 
integration contour $C$ is shown in Figure~\ref{fig:circle}, 
where the circle should be taken infinitely large in order 
to include all poles of the function $n(z)$.
Note that due to the fact that $\omega_n$
does not coincide with $\omega_l$ for any $n$ and $l$, the 
poles of $n(z)$ do not lie on the branch cuts of $C$.

We show now that the contribution of the large circle to the 
integral~\eqref{eq:sums_ab} vanishes. Because of the sum rule $\int_{-\infty}^{\infty}d\varepsilon
A(\bm p, \varepsilon)={\rm const}$, the quark
propagator for large $|z|$ has the scaling $D\propto z^{-1}$.
Therefore, for large $|z|$ the integrand in Equation~\eqref{eq:sums_ab} scales as $\propto {\tilde n}(z)z^{k-2}$ (recall that $f\propto z^k$). For the Fermi distribution function we have the asymptotics $n(z)\to_{{\rm Re}z\to \infty}0$ and $n(z)\to_{{\rm Re}z\to -\infty}1$.
Substituting $z=Re^{i\phi}$, $dz=iRe^{i\phi}d\phi$, and performing 
the limit $R\to\infty$, we can write for the integral along the circle 
\bea\label{eq:circle_int}
S_R\propto -\int_{C_R}\frac{dz}{2\pi i}z^{k-2}{\tilde n}(z)
=\frac{R^{k-1}}{4\pi}\left[\int_{-\pi/2}^{\pi/2}d\phi e^{i(k-1)\phi}-
\int_{\pi/2}^{3\pi/2}d\phi e^{i(k-1)\phi}\right]
\nonumber\\
=\frac{R^{k-1}}{\pi}
\left\{\begin{array}{ll} \frac{\sin (k-1)\pi/2}{k-1},\quad k\neq 1,\\
         0,\hspace{1.9cm} k= 1.
\end{array}\right.
\eea

As seen from Equation~\eqref{eq:circle_int}, the integral vanishes in the limit $R\to \infty$ if $k=0,1$.
If $k=2$, we have $S_R\propto R\to \infty$, which is 
purely real and can be dropped since in this case we need only the imaginary parts of $S[f]$ functions.
 As a result, we need to keep only the integrals along the two branch cuts shown in Figure~\ref{fig:circle}.
Then, using also Equation~\eqref{eq:spectralfunction0},
we obtain for Equation~\eqref{eq:sums_ab}
\bea\label{eq:residue2}
S[\hat{a},\hat{b}](\bm p, i\omega_n)=
\int_{-\infty}^{\infty} d\varepsilon  
{\tilde n}(\varepsilon)\Big\{f(\varepsilon-i\omega_n/2)
\Tr[\hat{a}_0 A(\bm p, \varepsilon)
\hat{b}_0 D(\bm p, \varepsilon -i\omega_n)]
\nonumber\\
+f(\varepsilon +i\omega_n/2){\Tr}
[\hat{a}_0 D(\bm p, \varepsilon+i\omega_n)
\hat{b}_0 A(\bm p, \varepsilon)]\Big\},
\eea
where we took into account that $n(\varepsilon-
i\omega_n)=n(\varepsilon)$.
Substituting the spectral 
representation \eqref{eq:propagator} into Equation~\eqref{eq:residue2}, changing the variables $\varepsilon\leftrightarrow\varepsilon'$ in the first term and performing 
analytical continuation via $i\omega_n\to \omega +i\delta$, we find
\bea\label{eq:residue4}
S[\hat{a},\hat{b}](\bm p, \omega)= \int_
{-\infty}^{\infty} \!\! d\varepsilon  \!\!\int_
{-\infty}^{\infty}\!\! d\varepsilon'~
\Tr[\hat{a}_0 A(\bm p, \varepsilon') 
\hat{b}_0 A(\bm p, \varepsilon)]
\frac{\tilde n(\varepsilon)f(\varepsilon+\omega/2)
-\tilde n(\varepsilon')f(\varepsilon'-\omega/2)}
{\varepsilon-\varepsilon' +\omega+i\delta}.
\eea
Substituting this expression into Equation~\eqref{eq:ring1_ab},
we obtain Equation~\eqref{eq:reloop} of the main text.

\pagebreak 
\reftitle{References}


\end{document}